\documentclass[twocolumn]{aastex61}
\usepackage{graphicx,mathtools}
\usepackage{amsmath,amssymb}
\usepackage{xparse}
\usepackage{xspace}
\usepackage{natbib}
\usepackage{listings}

\newcommand{\athena}{\texttt{Athena++}\xspace}

\newcommand\trho{\tilde{\rho}}
\newcommand\tT{\tilde{T}}
\newcommand\tP{\tilde{p}}

\newcommand{\mc}[1]{{#1}}

\DeclareMathOperator{\dd}{{\rm d}}

\newcommand{\der}[2]{\dfrac{\dd #1}{\dd #2}}
\NewDocumentCommand\pder{mmg}{\ensuremath{
		\IfNoValueTF{#3}
		{\dfrac{\partial #1}{\partial #2}}
		{\left(\dfrac{\partial #1}{\partial #2}\right)_{#3}}
}}
\begin{document}
	
\defcitealias{Chen19}{C19}

\title{An Extension of the Athena++ Framework for General Equations of State}
\shorttitle{Extension of Athena++ for General EOS}
\shortauthors{M. S. B. Coleman}
\keywords{equation of state; hydrodynamics; magnetohydrodynamics (MHD); methods: numerical}

\author{Matthew S. B. Coleman}
\affiliation{School of Natural Sciences, Institute for Advanced Study, Einstein Drive, Princeton, NJ 08540, USA; {\rm \url{mcoleman@ias.edu}}}

\begin{abstract}
We present modifications to the \athena framework to enable use of general equations of state (EOS).
Part of our motivation for doing so is to model transient astrophysics phenomena, as these types of events are often not well approximated by an ideal gas.
This necessitated changes to the Riemann solvers implemented in \athena. We discuss the adjustments made to the HLLC, and HLLD solvers and EOS calls required for arbitrary EOS.
We demonstrate the reliability of our code in a number of tests which utilize a relatively simple, but non-trivial EOS based on hydrogen ionization, appropriate for the transition from atomic to ionized hydrogen. Additionally, we perform tests using an electron-positron Helmholtz EOS, appropriate for regimes where nuclear statistical equilibrium is a good approximation. These new complex EOS tests overall show that our modifications to \athena accurately solve the Riemann problem with linear convergence \mc{and linear-wave tests with quadratic convergence}. We provide our test solutions as a means to check the accuracy of other hydrodynamic codes. Our tests and additions to \athena will enable further research into (magneto)hydrodynamic problems where realistic treatments of the EOS are required.
\end{abstract}

\section{Introduction}
While the assumption of an ideal gas is common in astrophysical fluid dynamics, there are many systems where this is not a valid approximation. 
This assumption is utilized due to its wide range of applicability and extremely simple equation of state (EOS); i.e. relations between fluid parameters including pressure, density and internal energy.
Supernovae \citep[e.g.][]{flash, chimera, sn_companion}, Kilonovae \citep[e.g.][]{2017ApJ...838L...2R}, and ionization instabilities in accretion disks \citep[e.g.][]{L01, H14, C16, S18, AMCVn} are just a few examples where a realistic treatment of the EOS is necessary, and the assumption of an ideal gas gives the wrong results.
These example also show the significance of realistic EOS in modeling transient events. As transient surveys such as ASASSN \citep{ASASSN1, ASASSN2}, Catalina \citep{Catalina1, Catalina2}, ZTF \citep{ZTF}, and the upcoming LSST \citep{LSST} continue to advance our observations of these events, we must also simultaneously increase the realism of our models. One key way of improving the fidelity of astrophysical fluid simulations is by utilizing realistic EOS.

For finite-difference (magneto)hydrodynamic, (M)HD, codes such as \texttt{Zeus} \citep{Zeus1, Zeus2}, it is relatively easy to use a general EOS \citep[as used by][]{H14}. However, for Godunov-type code such as \athena, it is much more challenging to incorporate a general EOS, as one must solve the Riemann problem using the same EOS (an aspect we will discuss in more detail later).
The seminal work of \citet{colella1985} enabled Godunov codes to use non-trivial EOS in a more robust manner, by addressing the Riemann problem for a general EOS. This paved the way for Godunov codes to adopt more realistic EOS, such as \texttt{FLASH} \citep{flash},
\texttt{CASTRO} \citep{castro},
and \texttt{Chimera} \citep{chimera}.
Here we extend the \athena framework \citep{athena} to perform non-relativistic (M)HD simulations using general EOS (subject to two assumptions discussed later). 
While \athena is capable of relativistic MHD \citep{White16}, the EOS extensions we describe here are only applicable to non-relativistic calculations. Although the methods described in this paper are not the first to incorporate realistic EOS in a MHD code, they make \athena\footnote{Available at \url{https://github.com/PrincetonUniversity/athena-public-version}} the first publicly-available open-source MHD code whose general EOS capabilities have been explicitly verified with a suite of (M)HD tests utilizing a non-trivial EOS. We note that \texttt{FLASH} \citep{flash} comes close to being able to make this claim, but they have not published any tests where they dynamically evolve a fluid utilizing a non-ideal EOS and verified against a known solution. 
Additionally, \texttt{CASTRO} \citep{castro} is a HD code which has previously run similar tests \citep{2015ApJS..216...31Z} but does not evolve magnetic fields.

To achieve the incorporation of general EOS we had to modify the Riemann solvers utilized by \athena. Accordingly, we summarize the Riemann problem in Section~\ref{sec:Riemann}. In Section~\ref{sec:methods} we describe the methods we use to solve the equations of (M)HD with a general EOS. We describe the generation of a series of tests to verify our code in Section~\ref{sec:tests} and analyze their results in Section~\ref{sec:dicussion}. In Section~\ref{sec:conclusions} we summarize our work and make some concluding remarks.

\section{The Riemann Problem}
\label{sec:Riemann}

The Riemann problem is a fundamental component of Godunov-type codes such as \athena, enabling them to evolve the fluid equations. In this section we describe the Riemann problem for the case of a 1D unmagnetized fluid described by the Euler equations:
\begin{align}
	\pder{\rho}{t}+\pder{}{x}(\rho v_x)&=0\\
	\pder{}{t}(\rho v_x)+\pder{}{x}(\rho v_x^2+p)&=0\\
	\pder{E}{t}+\pder{}{x}\left(E+p\right)&=0,
\end{align}
where $\rho$, $v_x$, $p$, and $e$ are, respectively, the mass density, speed, gas pressure, and internal energy density of the fluid in question. Additionally, $E=\rho v_x^2/2+e$ is the total energy density. A relation between $p$, $e$, and $\rho$ is required to close the system of equations which is provided by the EOS, e.g. 
$p=p(\rho, e),\text{ or }e=e\left(\rho, p\right)$,
which we assume to obey
\begin{align}
\left(\frac{\partial p}{\partial\rho}\right)_{e}&>0 \label{eqn:unique1}\\
\left(\frac{\partial p}{\partial e}\right)_{\rho}&>0  \label{eqn:unique2},
\end{align}
in the rest-frame of the fluid.
These assumptions are sufficient, but not necessary to guarantee that the Riemann problem has a unique solution \citep[see e.g.][hearafter \citetalias{Chen19}]{Chen19}.

The Riemann problem inquires how to describe the evolution of two semi-infinite constant fluid states,
\begin{align}
\mathbf{U}(x<0)=\mathbf{U}_{L}&=\left(\rho_L, v_{x,L}, p_L\right)\\
\mathbf{U}(x>0)=\mathbf{U}_{R}&=\left(\rho_R, v_{x,R}, p_R\right),
\end{align}
separated by a partition (at $x=0$) after its removal (at $t=0$). The majority of modern grid based hydrodynamic codes (including \athena) solve this problem at every cell interface at every time-step to compute the fluxes across these interfaces. Accordingly, it is necessary to revisit the Riemann problem when considering realistic EOS.

\begin{figure}
	\includegraphics[width=\linewidth]{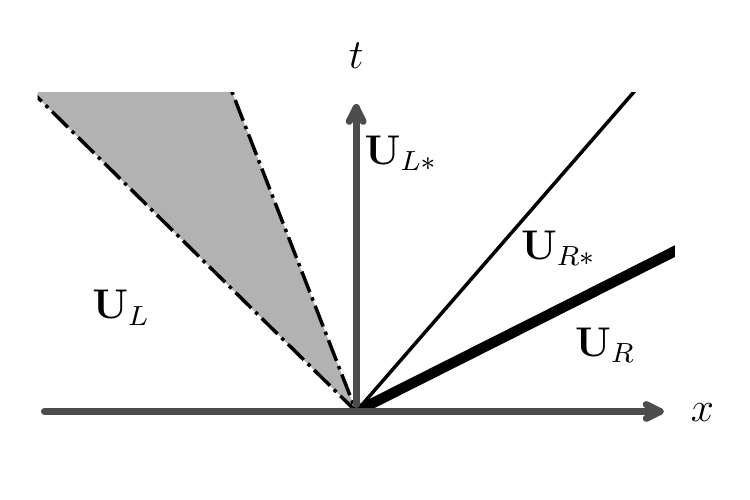}
	\vspace*{-3em}
	\caption{An example solution to an arbitrary Riemann problem illustrated as a Riemann fan. The horizontal axis $x$ is position, and the vertical axis is time. The four constant fluid states from left to right are $\mathbf{U}_{L}$, $\mathbf{U}_{L*}$, $\mathbf{U}_{R*}$, and $\mathbf{U}_{R}$. Here the left wave (shaded gray) is a rarefaction wave, middle is a contact discontinuity (thin-black line), and right is a shock wave (bold-black line).
	}
	\label{fig:fan}
\end{figure}

Often the solution to a Riemann problem is shown graphically as a ``Riemann fan" (see Fig.~\ref{fig:fan}). The Riemann fan visibly demonstrates that the solution is a function of only $x/t$, where $t$ is the time since the partition was removed and $x$ is the distance from the partition's location at $t=0$. It also shows that the solution consists of four different constant fluid states (from left to right: $\mathbf{U}_{L}$, $\mathbf{U}_{L*}$, $\mathbf{U}_{R*}$, and $\mathbf{U}_{R}$), separated by three waves. The outer two waves are either shock-waves or rarefaction-waves, while the middle wave is always a contact wave/discontinuity.

Recently, \citetalias{Chen19} closely examined the Riemann problem with particular care given to general EOS and demonstrate that it can be solved exactly to arbitrary precision. However, doing so is too computationally expensive for most practical purposes (even for an ideal gas). To make hydrodynamic simulation numerically tractable, (M)HD code tend to solve the Riemann problem approximately.

\citetalias{Chen19} also showed that generalizing the Riemann problem from an ideal gas to a more realistic EOS requires utilizing the precise definition of the adiabatic sound speed
\begin{align}
a^2\equiv\pder{p}{\rho}{s},
\end{align}
where the subscript $s$ denotes that the derivative is taken at constant specific entropy.

\section{Methods}
\label{sec:methods}

We start with the \athena code framework \citep{athena} which can solve the equations of MHD (here we neglect diffusive effects such as viscosity and resistivity)\footnote{Although, \athena is capable of including these effects.}:
\begin{align}
	\pder{\rho}{t}+\nabla\cdot\left(\rho\mathbf{v}\right)&=0\\
	\pder{\rho\mathbf{v}}{t}+\nabla\cdot\left[\rho\mathbf{vv}+\left(p+\dfrac{B^2}{2}\right)\mathbf{I}-\mathbf{BB}\right]&=0\\
	\pder{E}{t}+\nabla\cdot\left[\left(E+p\right)\mathbf{v}-\mathbf{B}\left(\mathbf{B}\cdot\mathbf{v}\right)\right]&=0\\
	\pder{\mathbf{B}}{t}-\nabla\times\left(\mathbf{v}\times\mathbf{B}\right)&=0,
\end{align}
where $\mathbf{v}$ is the fluid velocity, $\mathbf{I}$ is the identity tensor, $\mathbf{B}$ is the magnetic field, and $E=e+\rho v^2/2+B^2/2$ is the total energy density. For the entirety of this paper we use second-order piecewise-linear primitive\footnote{Characteristic reconstruction does not support general EOS.} reconstruction with the second-order van Leer time integrator.

\subsection{EOS Framework}
Here we describe the requirements and constraints of our general EOS framework. This framework here should be thought of as a means to implement a variety of EOS for use with \athena, with no particular application in mind.
As previously stated, we assume that EOS used in this framework obeys Eqns. \ref{eqn:unique1} and \ref{eqn:unique2}.
Implicit in this statement is that these derivatives are well defined and behaved.
The EOS must also provide the following functions for \athena to be able to evolve the fluid equations:
\begin{align}
\label{eqn:eos1}
p&=p\left(\rho, e\right)\\
\label{eqn:eos2}
e&=e\left(\rho, p\right)\\
\label{eqn:eos3}
a^2&=a^2\left(\rho, p\right)=\pder{p}{\rho}{s}.
\end{align}
The first two of these functions are used to convert primitives to conservatives and the inverse, respectively. \athena uses the total energy density ($E=e+\rho v^2/2+B^2/2$) as a conserved variable and thermal pressure $p$ as a primitive; accordingly, $e$ is computed from $E$ by subtracting off the other energy components before the conversion to pressure is made.
Finally, the adiabatic sound speed is used to determine the CFL condition and in the Riemann solver (e.g. for computing characteristic speeds and fluxes). Any EOS which is able to provide these functions and obey Eqns. \ref{eqn:unique1} and \ref{eqn:unique2} can work in this framework, even a piecewise EOS.

The generalization from an ideal gas to a more realistic EOS in \athena can be schematically represented as follows:
\begin{align}
\label{eqn:translate1}
p=\left(\gamma-1\right)e\;&\Rightarrow\;p=p\left(\rho, e\right)\\
\label{eqn:translate2}
e=\dfrac{p}{\left(\gamma-1\right)}\;&\Rightarrow\;e=e\left(\rho, p\right)\\
\label{eqn:translate3}
a^2=\gamma\dfrac{p}{\rho}\;&\Rightarrow\;a^2=a^2\left(\rho, p\right),
\end{align}
where $\gamma$ is the adiabatic index for the ideal gas.  All of these translations are intuitive, however it is non-trivial to show that these translations preserve all the characteristic speeds associated with the Riemann problem \citepalias[see Appendix B of][]{Chen19}. With one caveat (see Appendix~\ref{sec:wavespeeds}), these three translations cover all the changes required to generalize \athena for realistic EOS. In short, the changes we made to \athena can be summarized as follows: we found all instances of $\gamma$ and replaced them using one of the above translations.

\subsection{Tabular EOS}
\label{sec:eos-table}

While Eqns.~\ref{eqn:eos1}-\ref{eqn:eos3} can written in a (semi-)analytic fashion, they may also be implemented with interpolated tables. Here we present the details of the tabular EOS utility that we have created for use with our general EOS framework, although the framework is more flexible and extendible, if the need for more complicated tables arises, such as the Helmholtz EOS \citep{helmeos} discussed in Section~\ref{sec:helm_tests}.

The functions~\ref{eqn:eos1}-\ref{eqn:eos3} need to be precomputed in some discretized way. In stead of precomputing these functions directly, we tabulate\footnote{We use the convention that $\log q\equiv\log_{10}q$ and $\ln q\equiv\log_eq$.}
\begin{align}
\label{eqn:interp1}
\log\left(p/e\right)&\left(\log\rho, \log e + \eta \log \rho\right)\\
\label{eqn:interp2}
\log\left(e/p\right)&\left(\log\rho, \log p + \eta \log \rho\right)\\
\label{eqn:interp3}
\log\left(\Gamma_1\right)&\left(\log\rho, \log p + \eta \log \rho\right),
\end{align}
where $\eta$ is a user specified constant, and
\begin{align}
\Gamma_1\equiv\pder{\ln p}{\ln\rho}{s}=\dfrac{\rho}{p}a^2.
\end{align}
These all take the form of $\log\left(q\right)\left(\log\rho, \log \epsilon \right)$, where $q$ is a dimensionless quantity and $\epsilon$ is either $e\rho^\eta$ or $p\rho^\eta$ (which both have the same dimensionality).
For simplicity, we discretize these functions with a regularly spaced rectangular grid:
\begin{align}
\delta\log\rho&
\equiv\log\rho_{i+1}-\log\rho_i
=\dfrac{\log\rho_{N_\rho-1}-\log\rho_0}{N_\rho-1}\\
\delta\log \epsilon&
\equiv\log \epsilon_{j+1}-\log \epsilon_j
=\dfrac{\log \epsilon_{N_\epsilon-1}-\log \epsilon_0}{N_\epsilon-1},
\end{align}
where $N_\rho$ and $N_\epsilon$ are the sizes of the table in the $\rho$ and $\epsilon$ directions respectively,  $i\in\left\lbrace 0..(N_\rho-1) \right\rbrace$, and $j\in\left\lbrace 0..(N_\epsilon-1) \right\rbrace$. We allow for an arbitrary shift between pressure and internal energy,
\begin{align}
e_j=c_0 p_j\;\;\forall j\in\left\lbrace 0..(N_\epsilon-1) \right\rbrace
\end{align}
with $c_0$ an arbitrary constant (set to one by default), but $e$ and $p$ are both required to have the same number of points ($N_\epsilon$). 
Bilinear interpolation is used to estimate the values of \ref{eqn:interp1}-\ref{eqn:interp3} in-between the discretized points; this guarantees the preservation of monotonicity. If data is requested from off the table, then linear extrapolation is used, although it is best to prevent this from happening by creating sufficiently large tables and setting floors for $p$, $e$ and $\rho$ wisely. An alternative that is feasible in this framework (although not currently implemented) is to use a different EOS outside the table domain.

One of the motivations for choosing \ref{eqn:interp1}-\ref{eqn:interp3} as the tabulated quantities, is that for an ideal gas these are constants, i.e.
\begin{align}
\dfrac{p}{e}&=\gamma-1\\
\dfrac{e}{p}&=\dfrac{1}{\gamma-1}\\
\Gamma_1&=\gamma.
\end{align}
This enables a table with $N_\rho\geq2$ and $N_\epsilon\geq2$ to exactly (to machine precision) reproduce the results of explicitly using the ideal gas EOS, which we have verified with \athena.

\subsection{Riemann Solvers}

For HD problems we utilize the HLLC \citep{toro1994} Riemann solver. Before the work presented here, \athena used the Roe average to approximate the middle state as a means to estimate the extremal wave-speeds in the HLLC solver \citep[see e.g. Sections 10.5.1 and 11.3.3 of][]{ToroBook}.
After experimenting with a few wave-speed estimators \citep[including that of][]{HU09} we discovered that the PVRS (primitive variable Riemann solver) method described in Sections 9.3 and 10.5.2 of \citet{ToroBook} resulted in significant reduction of errors and improved convergence in tests 3 and 4 (see Section~\ref{sec:tests} and Table~\ref{tab:tests}), without negatively affecting the accuracy of solutions using an ideal gas EOS. Accordingly, we now use this method to estimate the extremal wave-speeds. The details of these two wave-speed estimators are given in Appendix~\ref{sec:wavespeeds}. While for MHD problems we use the HLLD Riemann solver with the wave-speed estimator given by Equation 12 of \citet{HLLD}, as originally used in \athena.

\section{Generating Tests}
\label{sec:tests}

The preexisting literature on (M)HD tests involving a non-trivial EOS is rather sparse, most of which is not done with astrophysical contexts in mind \citep[e.g.][]{HU09, doi:10.1063/1.4851415}. Only \citetalias{Chen19} and \citet{2015ApJS..216...31Z} generate exact solutions for hydrodynamic tests involving a non-trivial EOS in astrophysical contexts. Additionally, none of these sources publishes their exact solutions, and some do not even generate exact solutions in the first place \citep[e.g.][]{doi:10.1063/1.4851415}, preventing \mc{error comparison between different techniques. We did run the three variants of the ``DG1" tests from \citet{doi:10.1063/1.4851415} and recovered comparable results. We did not attempt to run the tests of \citet{HU09} and \citet{2015ApJS..216...31Z}, as these would require our EOS functions Eqns.~\ref{eqn:eos1}-\ref{eqn:eos3} to take additional arguments such as passive scalars, which is beyond the scope of this paper.}
As this work was developed in conjunction with \citetalias{Chen19}, 
we developed tests independently 
and we wanted to create a series of tests based on an analytic EOS (not done by \citealt{2015ApJS..216...31Z}), requiring us 
to develop our own series of Riemann problem tests to validate our code.


\subsection{Hydrogen HD Riemann Tests}
We use a relatively simple EOS which consists of only electrons, neutral hydrogen, and ionized hydrogen, with an ionization fraction $x$ governed by Saha's equation; see Appendix~\ref{sec:eos} for more details. This EOS is designed to test the codes ability to accurately evolve fluids through ionization transitions (as is necessary for ionization instabilities in accretion disks; see e.g. \citealp{H14,AMCVn}). We use an arbitrary-precision general EOS Riemann Solver based on that developed by \citetalias{Chen19}\footnote{We utilize different bracketing bounds for root finding, which does not affect the outcome of the solution.} to generate solutions for comparison with \athena. All tests using this hydrogen EOS (tests 1-7) are done with the publicly available version 19.0 of \athena.

When running these tests within \athena, every time an EOS call is made, first the code performs a root find\footnote{We use the Brent–Dekker method to compute the temperature to a precision of one part in $10^{12}$ and our bounding guesses are computed assuming the ionization fraction is either one or zero.} to determine the temperature, which is used along with density to analytically compute the required EOS quantity. For tests, this is substantially better than using a lookup table, as the convergence would be sensitive to the details and resolution of the table implementation.

\begin{deluxetable}{cccc}
	\tablecaption{Assumed Units\label{tab:units}}
	\tablehead{
		\colhead{Quntity} & \colhead{Symbol} & \colhead{Expression} & \colhead{cgs value}
	}
	\startdata
	mass & $m_p$ & $m_p$ & 1.6726219e-24\\
	temperature & $T_{\rm ion}$ & $\dfrac{1}{k}\dfrac{\alpha^2m_ec^2}{2}$ & 157,888\\
	number density & $n_{\rm q}$ & $\left(\dfrac{2 \pi  m_e k T_{\rm ion}}{h^2}\right)^{3/2}$ & 1.514892e23\\
	density & $\rho_{\rm u}$ & $m_p n_{\rm q}$ & 0.253384\\
	pressure & $P_{\rm u}$ & $n_{\rm q} k T_{\rm ion}$ & 3.302272e12\\
	magnetic field & $B_{\rm u}$ & $\sqrt{P_{\rm u}}$ & 1.8172154e6\\
	speed & $v_{\rm u}$ & $\sqrt{k T_{\rm ion} / m_p}$ & 3.6100785e6\\
	length & $\ell_{\rm u}$ & $n_{\rm q}^{-1/3}$ & 1.8758844e-8\\
	time & $t_{\rm u}$ & $\ell_{\rm u} / v_{\rm u}$ & 5.196243e-15\\
	\enddata
	\tablecomments{These units are chosen for convienence of calculations with our EOS (see Appendix~\ref{sec:eos}) and the subscript ``u" stands for unit.}
\end{deluxetable}

\begin{deluxetable*}{ccccccccc}
	\tablecaption{Hydrogen HD Riemann Tests\label{tab:tests}}
	\tablehead{
		\colhead{Test \#} & \colhead{Test Type} & \colhead{$\rho_{\rm l}$} & \colhead{$v_{\rm l}$} & \colhead{$T_{\rm l}$} & \colhead{$\rho_{\rm r}$} & \colhead{$v_{\rm r}$} & \colhead{$T_{\rm r}$} & \colhead{$\Delta t/\Delta x$}
	}
	\startdata
1 & Sod-like & 1e-07 & 0.0 & 0.15 & 1.25e-08 & 0.0 & 0.062 & 0.25\\
2 & Sod-like & 4e-06 & 0.0 & 0.12 & 4e-08 & 0.0 & 0.019 & 0.3\\
3 & Asym. Shock-Shock & 8e-07 & 1.1 & 0.006 & 4e-07 & -1.7 & 0.006 & 1.5\\
4 & Asym. Shock-Shock & 5e-07 & 1.5 & 0.006 & 4e-07 & -1.8 & 0.006 & 1.5\\
5 & Sym. Rare-Rare & 8e-05 & -0.8 & 0.095 & 8e-05 & 0.8 & 0.095 & 0.25\\
6 & Asym. Rare-Rare & 6e-05 & -0.5 & 0.095 & 8e-05 & 0.9 & 0.095 & 0.25\\
	\enddata
	\tablecomments{The left (l) and right (r) density ($\rho$), speed ($u$), and temperature ($T$), as well as $\Delta t/\Delta x$ are given in the units listed in Table~\ref{tab:units}.}
\end{deluxetable*}

\begin{deluxetable*}{ccccccc|ccccccc}
	\tablecaption{Test 7 (MHD Riemann Problem) Inital Conditions\label{tab:rj2a}}
	\tablehead{
		\colhead{$\rho_{\rm L}$} & \colhead{$v_{x,{\rm L}}$} & \colhead{$v_{y,{\rm L}}$} & \colhead{$v_{z,{\rm L}}$} & \colhead{$B_{y,{\rm L}}$} & \colhead{$B_{z,{\rm L}}$} & \colhead{$p_{\rm L}$} &
		\colhead{$\rho_{\rm R}$} & \colhead{$v_{x,{\rm R}}$} & \colhead{$v_{y,{\rm R}}$} & \colhead{$v_{z,{\rm R}}$} & \colhead{$B_{y,{\rm R}}$} & \colhead{$B_{z,{\rm R}}$} & \colhead{$p_{\rm R}$}
	}
	\startdata
	1.08 & 1.2 & 0.01 & 0.5 & $3.6/\sqrt{4\pi}$ & $2/\sqrt{4\pi}$ & 0.95 &
	1 & 0 & 0 & 0 & $4/\sqrt{4\pi}$ & $2/\sqrt{4\pi}$ & 1\\
	\enddata
	\tabletypesize{\normalsize}
	\tablecomments{Initial conditions for test 7, based on ``test 2a" from \citet{RJ95}. $B_x=2/\sqrt{4\pi}$ is constant throughout the simulation domain. This test is run for $\Delta t/\Delta x=0.2$. Density is in units of $10^{-7}\rho_{\rm u}$, velocity in $\sqrt{0.2}\,v_{\rm u}$, magnetic fields in $\sqrt{2\times 10^{-8}} \,B_{\rm u}$, pressure in $2\times 10^{-8} \,p_{\rm u}$ (see Table~\ref{tab:units} for unit deffinitions).}
	\label{tab:mhd-ic}
\end{deluxetable*}

\begin{deluxetable*}{ccccccccc}
	\tablecaption{Helmholtz HD Riemann Tests\label{tab:helm_tests}}
	\tablehead{
		\colhead{Test \#} & \colhead{Test Type} &
		\colhead{$\rho_{\rm l}$} & 
		\colhead{$\dfrac{v_{\rm l}}{10^8}$} & 
		\colhead{$\dfrac{p_{\rm l}}{10^{16}}$} & 
		\colhead{$\rho_{\rm r}$} & 
		\colhead{$\dfrac{v_{\rm r}}{10^8}$} & 
		\colhead{$\dfrac{p_{\rm r}}{10^{16}}$} & 
		\colhead{$\dfrac{\Delta t/\Delta x}{10^{-8}}$}
		\vspace*{.5em}
	}
	\startdata
10 & Sod-like & 1.0 & 0.0 & 0.7 & 0.125 & 0.0 & 0.015 & 0.25\\
11 & Asym. Shock-Shock & 0.8 & 1.1 & 0.05 & 0.4 & -1.7 & 0.03 & 0.6\\
	\enddata
	\tablecomments{The left (l) and right (r) density ($\rho$), speed ($u$), and pressure ($p$), as well as $\Delta t/\Delta x$ are given in cgs units. For these tests we also set $\bar{A}=\bar{Z}=1$ \citep[see][for more details]{helmeos}.}
\end{deluxetable*}

Due to the nature of our EOS, a set of units naturally arises (see Table~\ref{tab:units}). Unless otherwise specified, all quantities are given in these units. Two notable exceptions are the length and duration of the simulation runs, $\Delta x$ and $\Delta t$ respectively. This is because Riemann solutions are scale free, depending only on $x/t$. Therefore we only specify the ratio of $\Delta t/ \Delta x$ (in units of $1/v_{\rm u}$).

One trivial test that we ran was a 1D sinusoidal-linear-wave at $\rho=p=1$ (making $\Gamma_1=1.615$) with a wave amplitude of $\delta\rho=10^{-6}$. The wave was initialized in an eigenmode with the sound speed as the expected propagation speed. We verified that the wave returned to its original position after one sound-crossing time. 

As this test has no way of probing the accuracy of non-linear HD, we also define 6 different Riemann problems, 
listed in Table~\ref{tab:tests},
to test the accuracy and convergence of \athena in non-linear HD. The simulation domain spans $\pm\Delta x/2$ and are run for a duration of $\Delta t$. The initial conditions are two separate constant (left/right) states with a discontinuity at $x=0$. We specify the density ($\rho$), speed ($v_x$), and temperature ($T$) for the left and right initial states. Traditionally, pressure ($p$) is used instead of temperature, however Eqn.~\ref{eqn:p} allows us to readily compute $p(\rho, T)$, and using temperature makes it easier to define problems where the ionization state changes significantly. The corresponding pressures of these initial states can be found in Appendix~\ref{sec:sol}.

To classify the errors we define the $L^1$ and $L^2$ norms as
\begin{align}
L^1\left(f\right)&=\dfrac{1}{N}\sum_i \left|\Delta f_i\right|\\
L^2\left(f\right)&=\dfrac{1}{N}\sqrt{\sum_i \left|\Delta f_i\right|^2},
\end{align}
where $N$ is the number of cells used in a simulation.
For $f\ne x,t$, we also define the cell-wise error between the simulation data and the exact solution
\begin{equation}\label{key}
\Delta f \equiv f_{\athena} - f_{\rm exact}.
\end{equation}

As the general Riemann problem contains discontinuities, the expected convergence is linear\mc{\footnote{\mc{In actuality sub-linear convergence is the expected asymptotic behavior for contact discontinuities and other linear discontinuities \citep{Banks_Aslam_Rider_2008}, but it is likely that we do not achieve high enough resolution for these type of errors to be dominant in most of our tests.}}}, i.e.  $L^1\left(f\right)\propto1/N$, and $L^2\left(f\right)\propto1/N$, where N is the number of cells. To test the convergence, each test is run with $N=64, 128, 256, 512, 1024, 2048$ cells. Additionally, we set the CFL number to $0.4$. We provide all test solutions in Appendix~\ref{sec:sol} to enable tests with other (M)HD codes. 

We first note the similarity between the tests presented here and those in \citetalias{Chen19}; as noted before, these works were developed simultaneously.
All of these tests are designed to exhibit substantial variations in both the ionization fraction $x$ and effective adiabatic index $\Gamma_1$, to test the code's ability to accurately describe non-ideal EOS.
The first two tests (1 and 2) are based the classical \citet{Sod} shock tube, with the initial left/right states at zero velocity and higher pressure on the left which drives a rightwards shock. Tests 3 and 4 are asymmetric double shocks, where both initial states have a supersonic inward flow. As we will discuss in Section~\ref{sec:dicussion} these prove to be the most changing tests. Tests 5 and 6 are respectively, symmetric and asymmetric double rarefaction wave tests.

\subsection{Hydrogen MHD Riemann Test}
\label{sec:mhd-tests}

In addition to the previous HD tests, it is also important to test that our code can accurately reproduce a known MHD problem with a general EOS.
Unfortunately, prior to this work, there were no published tests that have a known exact solution for MHD with a non-trivial EOS. \citet{doi:10.1063/1.4851415} do have MHD ``tests" of the van der Waals EOS, but they do not compare their numerical results to a known solution, and therefore \mc{it is impossible to say if our method is more accurate}. Despite this, we ran all three of their ``DG1" tests and visually compare the density and got approximately the same result.

This lack of proper tests required us to develop our own MHD tests.
The general MHD Riemann problem with a general EOS exactly is non-trivial and beyond the scope of this work. Instead, we restrain ourselves to the subset of MHD Riemann problems \citep[see e.g.][for details on the MHD Riemann problem]{RJ95} where all seven waves are discontinuities. To solve this type of Riemann problem, one only needs to consider the jump conditions across the discontinuities \citep{RJ95}:
\begin{align}
F_i\left[1/\rho\right]&=-\left[v_{x}\right]\\
F_i\left[v_{x}\right]&=\left[p_{\rm tot}-B_{x}^{2}\right]\\
F_i\left[v_{y}\right]&=-B_{x}\left[B_{y}\right]\\
F_i\left[v_{z}\right]&=-B_{x}\left[B_{z}\right]\\
F_i\left[B_{y}/\rho\right]&=-B_{x}\left[v_{y}\right]\\
F_i\left[B_{z}/\rho\right]&=-B_{x}\left[v_{z}\right]\\
F_i\left[E/\rho\right]&=\left[v_{x} p_{\rm tot}\right]-B_{x}\left[B_{x} v_{x}+B_{y} v_{y}+B_{z} v_{z}\right],
\end{align}
where $E=e+(\rho v^2 + B^2)/2$, $p_{\rm tot}=p+B^2/2$, $\left[q\right]=q_{i+1}-q_i$, and $F_i$ is the mass flux across the discontinuity separating the $i$ and $i+1$ states. This gives 49 equations (the seven above equations for each of the seven discontinuities) and 49 unknowns: the fluid parameters ($\rho_i,\,v_{x,i},\,v_{y,i},\,v_{z,i},\,B_{y,i},\,B_{z,i},\,p_i$) for the six intermediate states ($i\in\left\lbrace 2..7 \right\rbrace$) and the fluxes ($F_i$) across all seven discontinuities. 

To create our seventh test we utilize the initial left ($i=1$) and right ($i=8$) states used in ``test 2a" of \citet{RJ95} (see Table~\ref{tab:mhd-ic}). The difference that we introduce is the hydrogen EOS (Appendix \ref{sec:eos}), where \citet{RJ95} use an ideal EOS, and we tune the simulation units to achieve a large variation in $\Gamma_1$. We used the ideal gas solution presented in \citet{RJ95} as our initial guess for numerically determining the solution to the system of the above 49 equations to generate our test solution (Table~\ref{tab:sol-7}).

\subsection{Hydrogen Linear Wave Tests}
\label{sec:lin-wave_test}

\begin{figure*}
	\begin{center}
		\includegraphics[width=.5\linewidth]{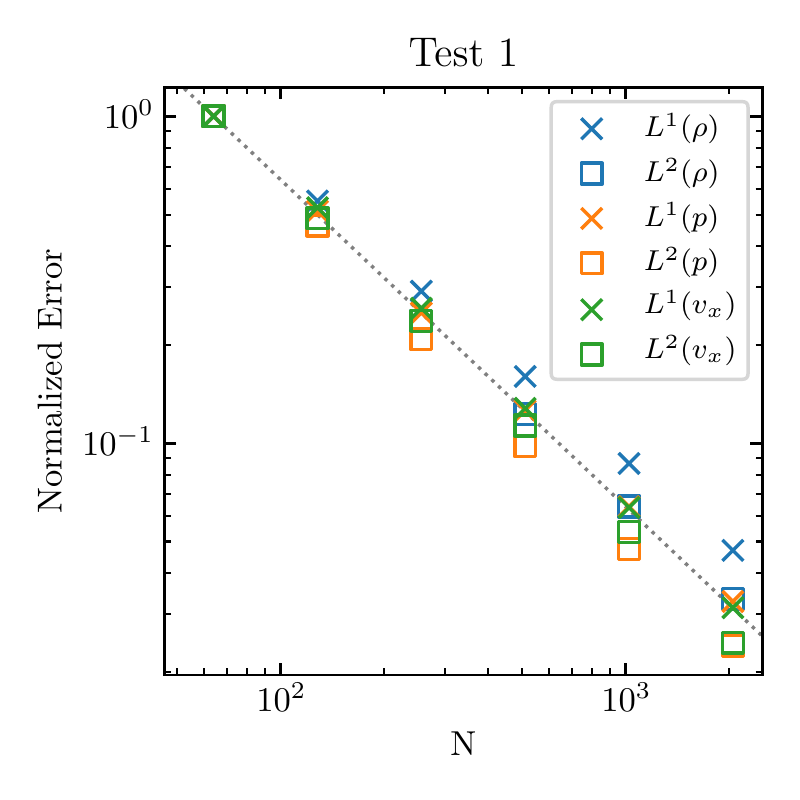}\\
		\includegraphics[width=.49\linewidth]{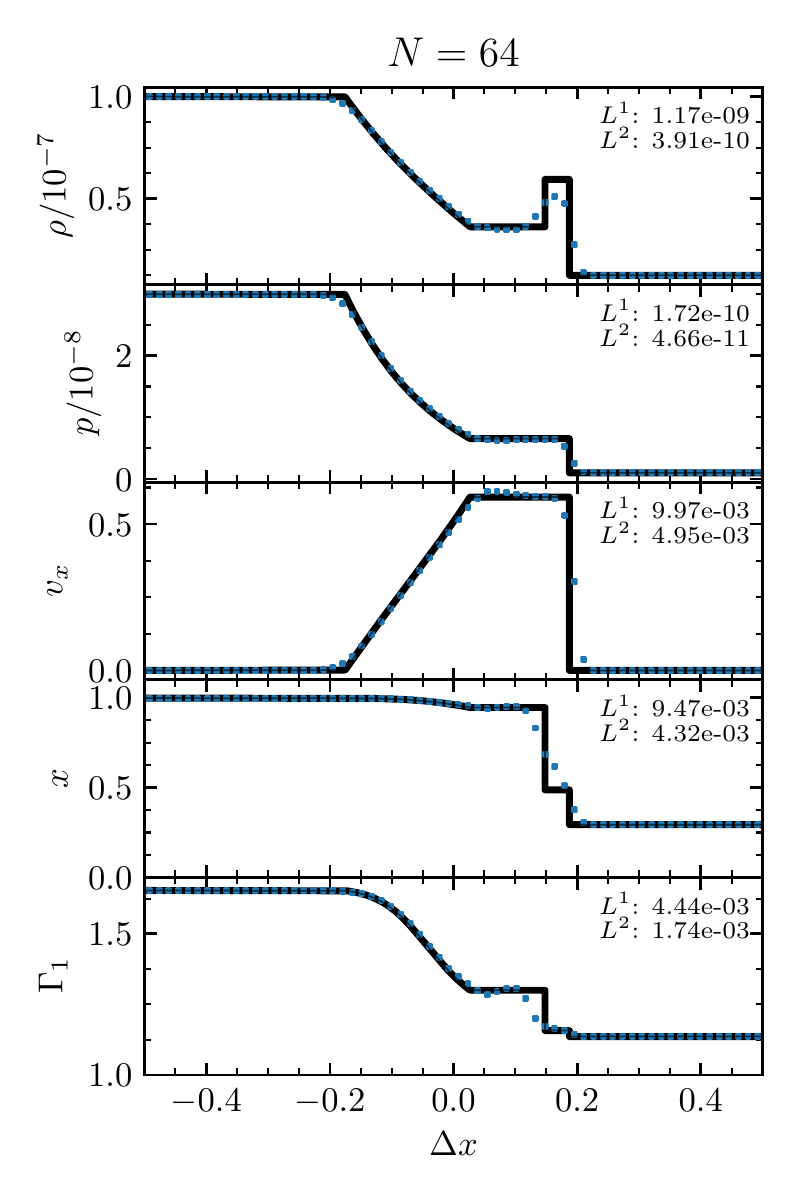}
		\includegraphics[width=.49\linewidth]{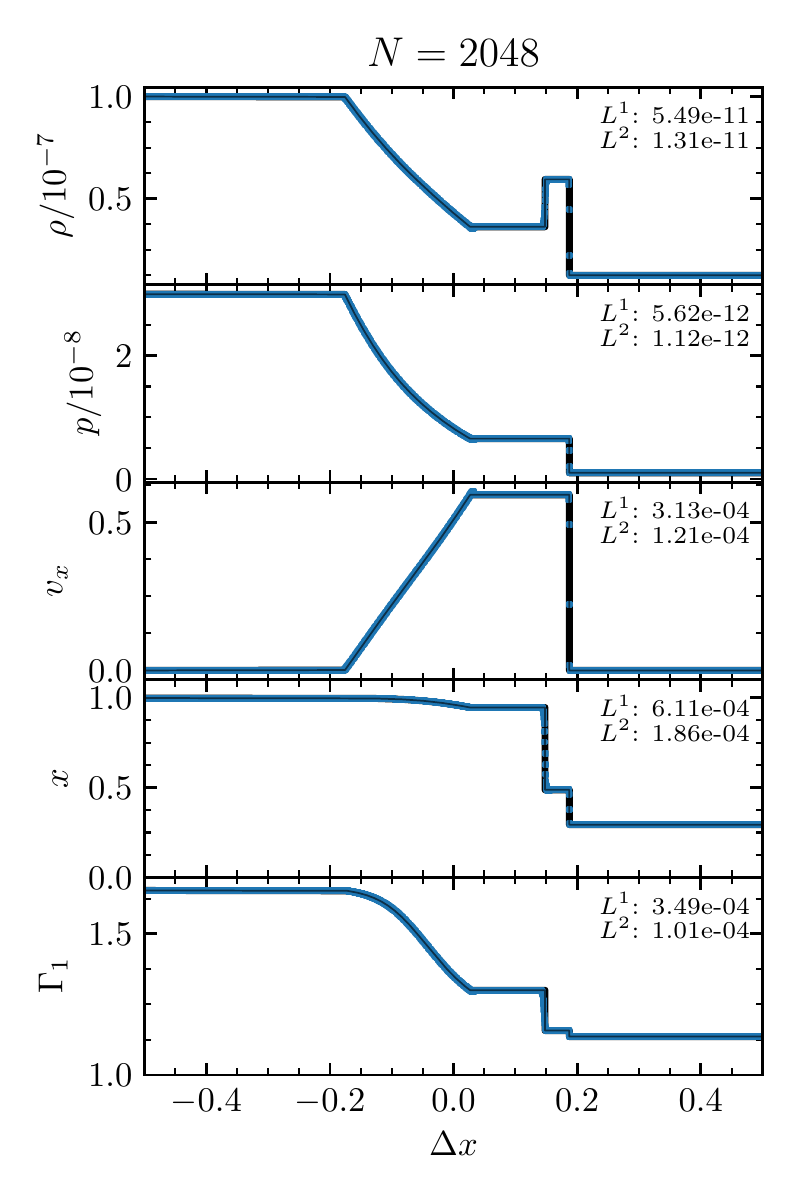}
	\end{center}
	\caption{Results for Riemann test 1 (see Table~\ref{tab:tests}). Top: $L^1,\,L^2$ errors for density, pressure and velocity, as a function of number of cells ($N$) normalized by their value at $N=64$. The dotted gray line indicates a linear trend. Bottom: Profiles of \athena results (blue points) and exact solution (black line) at $t=\Delta t$ for $N=64$ (left) and $N=2048$ (right).
		\vspace{1in}
	}
	\label{fig:test01}
\end{figure*}

\begin{figure*}
	\begin{center}
		\includegraphics[width=.5\linewidth]{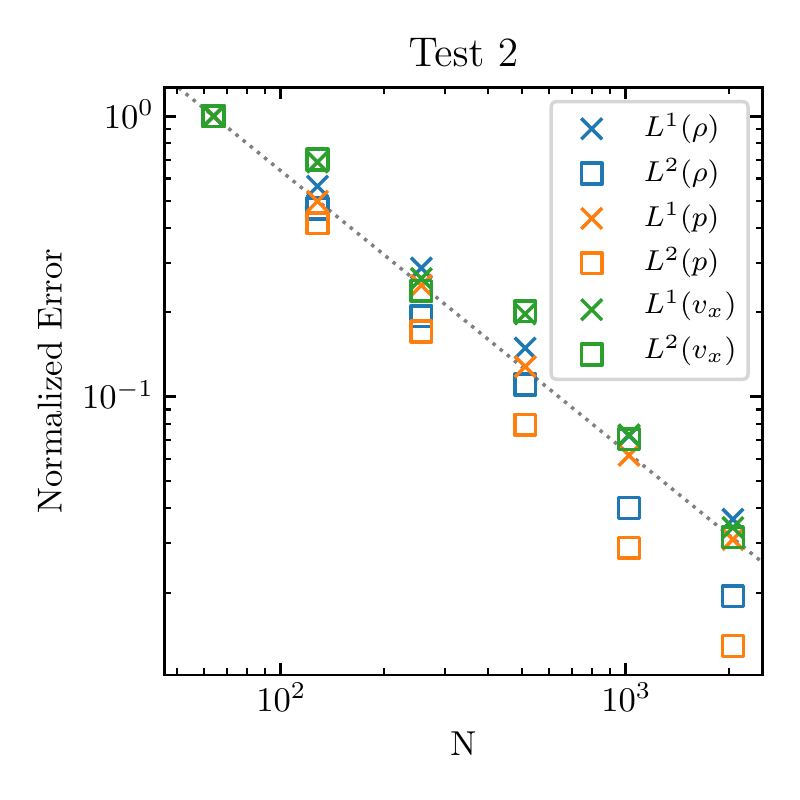}\\
		\includegraphics[width=.49\linewidth]{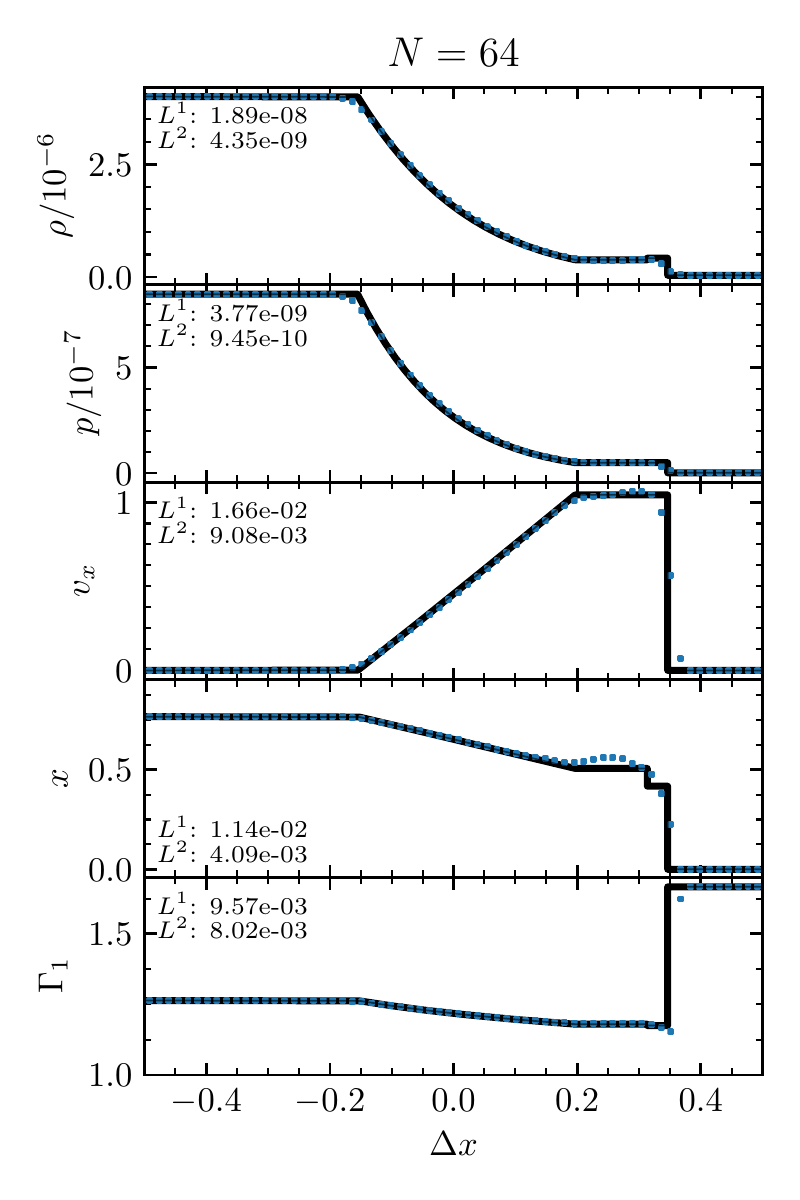}
		\includegraphics[width=.49\linewidth]{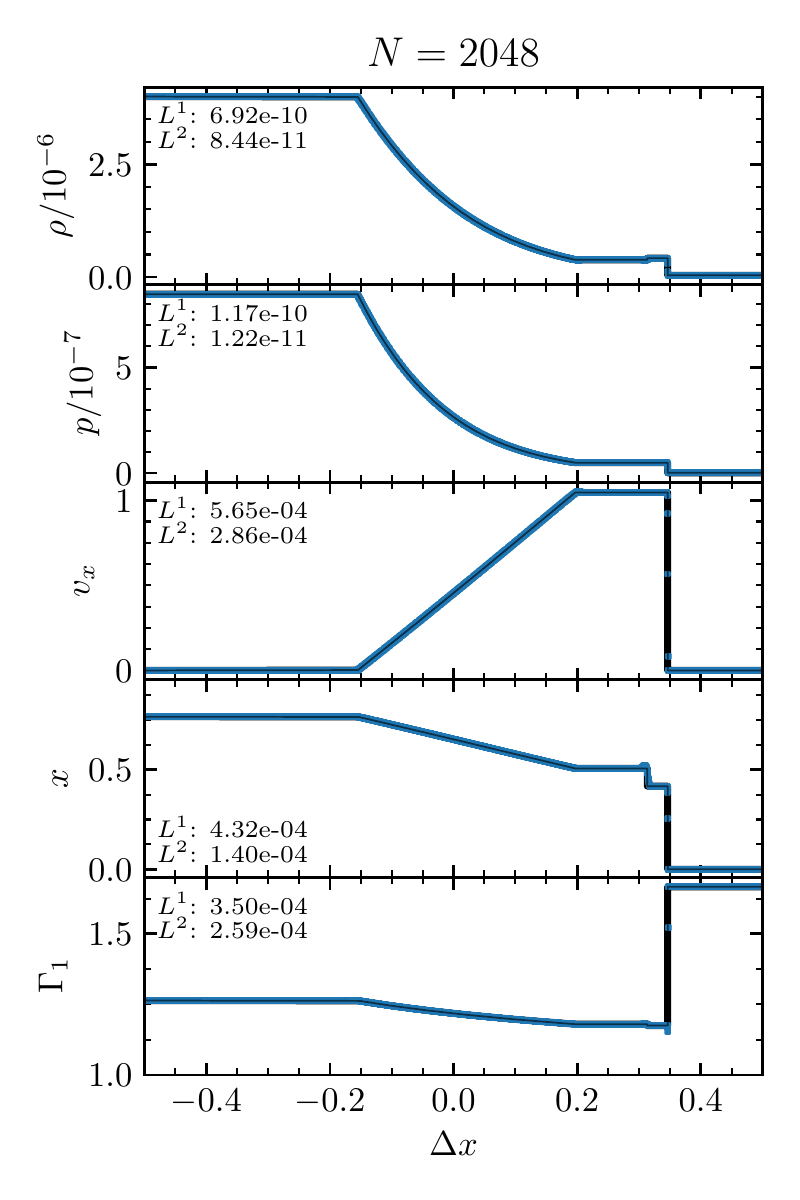}
	\end{center}
	\caption{Same as Fig.~\ref{fig:test01} but for Riemann test 2 (see Table~\ref{tab:tests}).
		\vspace{1in}
	}
	\label{fig:test02}
\end{figure*}

\begin{figure*}
	\begin{center}
		\includegraphics[width=.5\linewidth]{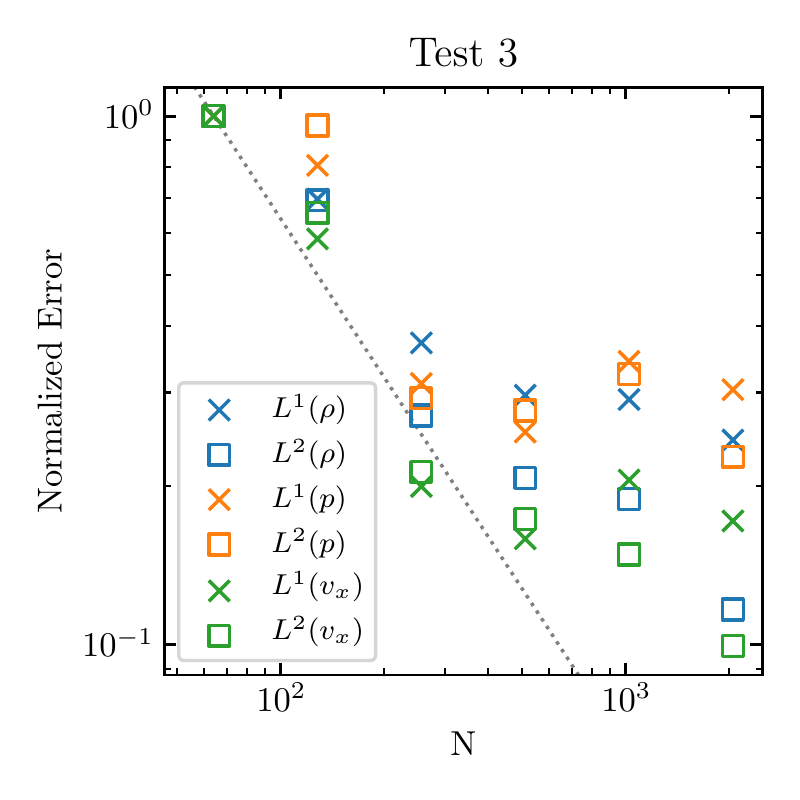}\\
		\includegraphics[width=.49\linewidth]{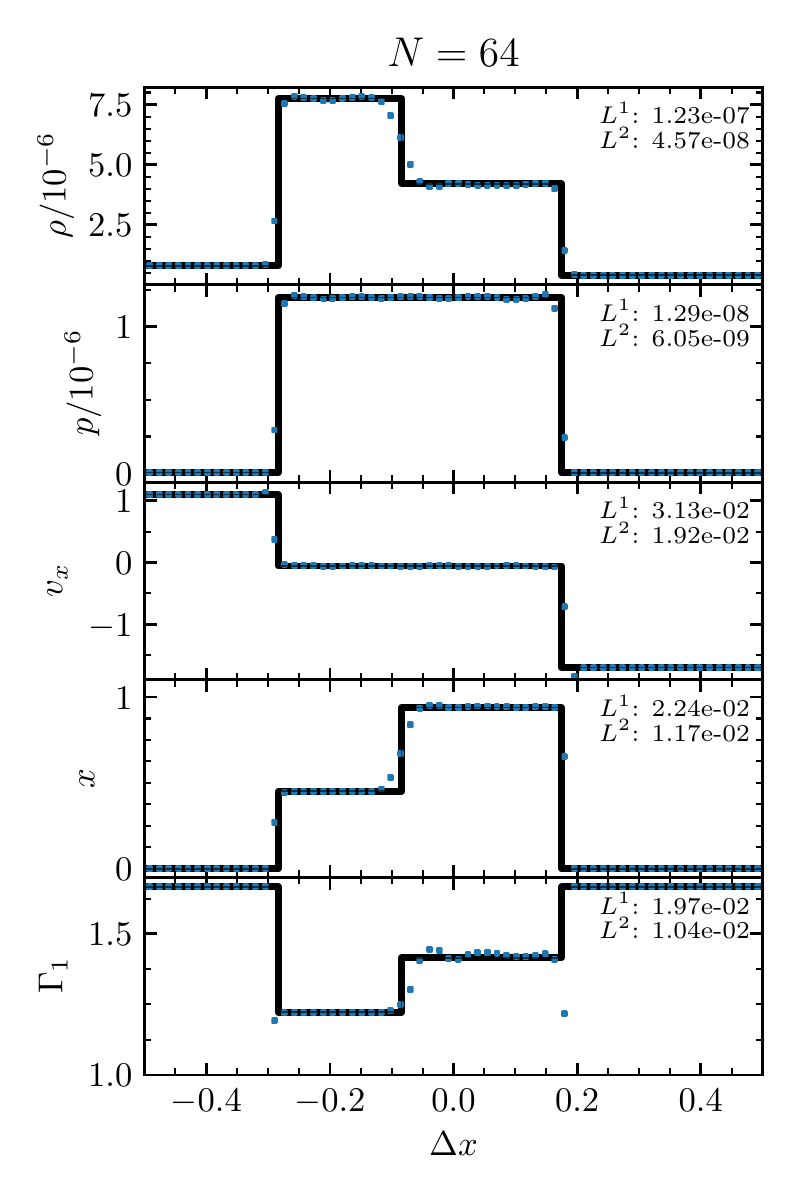}
		\includegraphics[width=.49\linewidth]{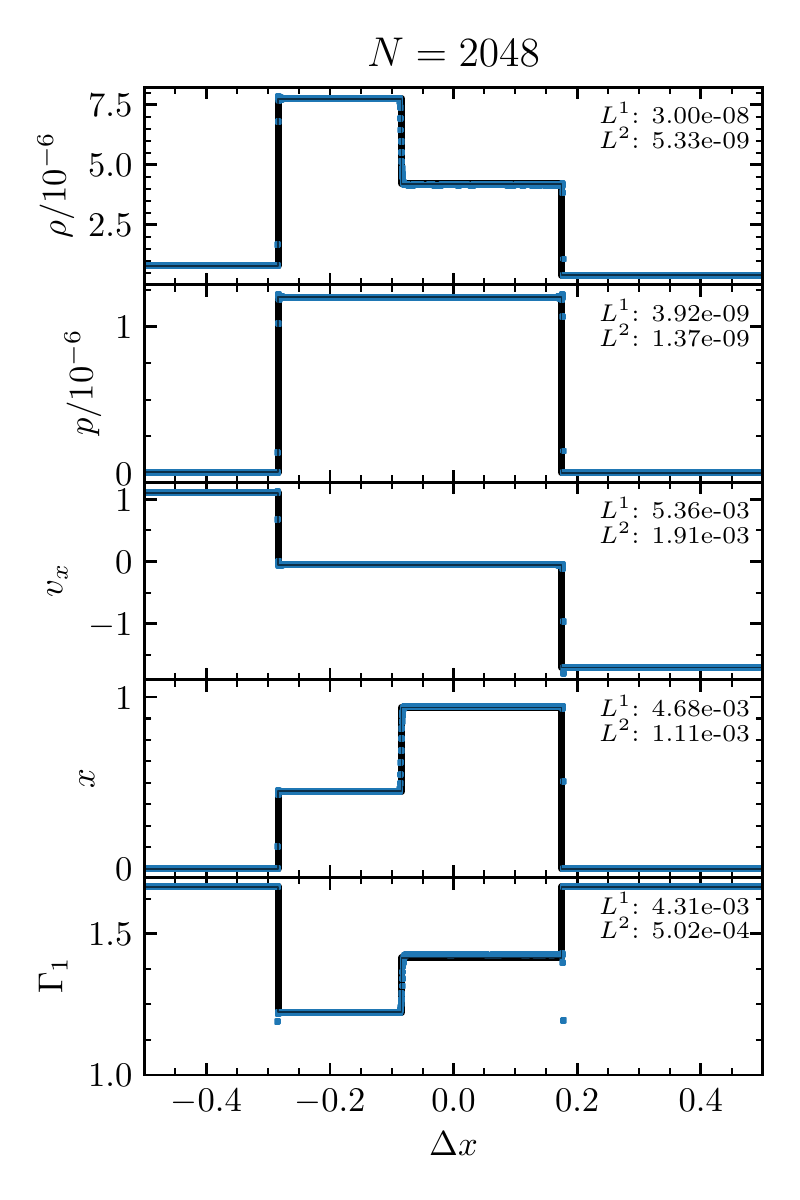}
	\end{center}
	\caption{Same as Fig.~\ref{fig:test01} but for Riemann test 3 (see Table~\ref{tab:tests}).
		\vspace{1in}
	}
	\label{fig:test03}
\end{figure*}

\begin{figure*}
	\begin{center}
		\includegraphics[width=.5\linewidth]{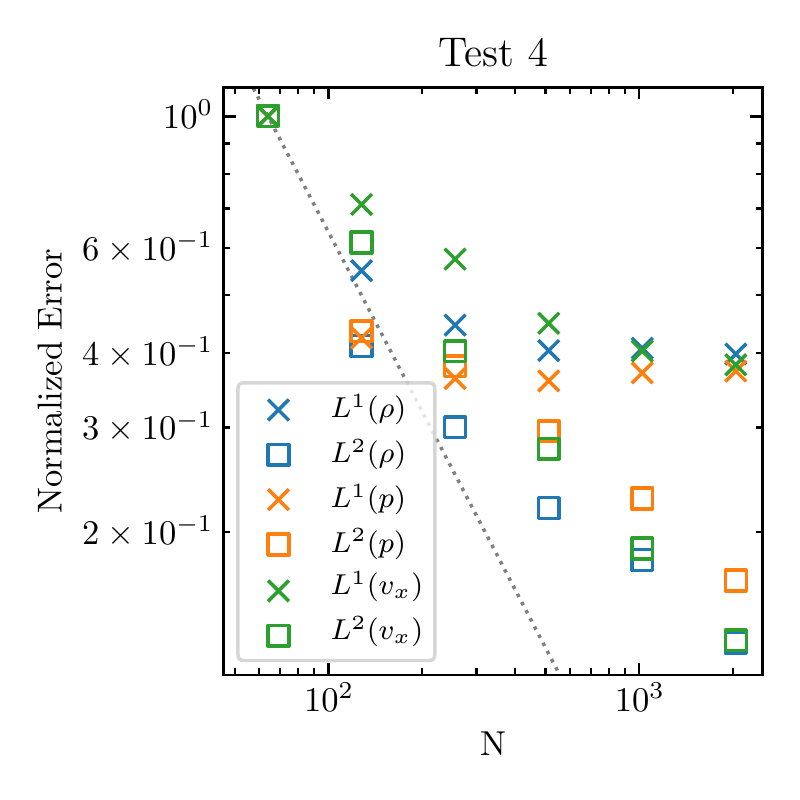}\\
		\includegraphics[width=.49\linewidth]{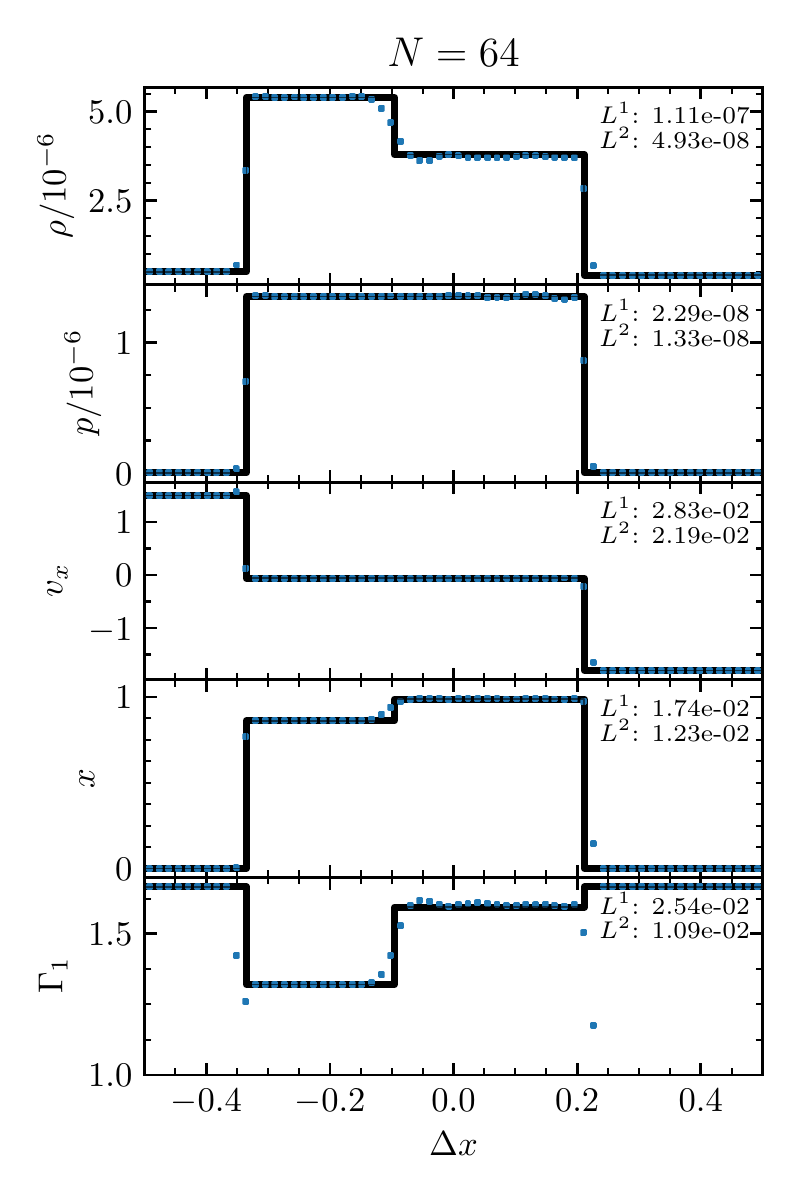}
		\includegraphics[width=.49\linewidth]{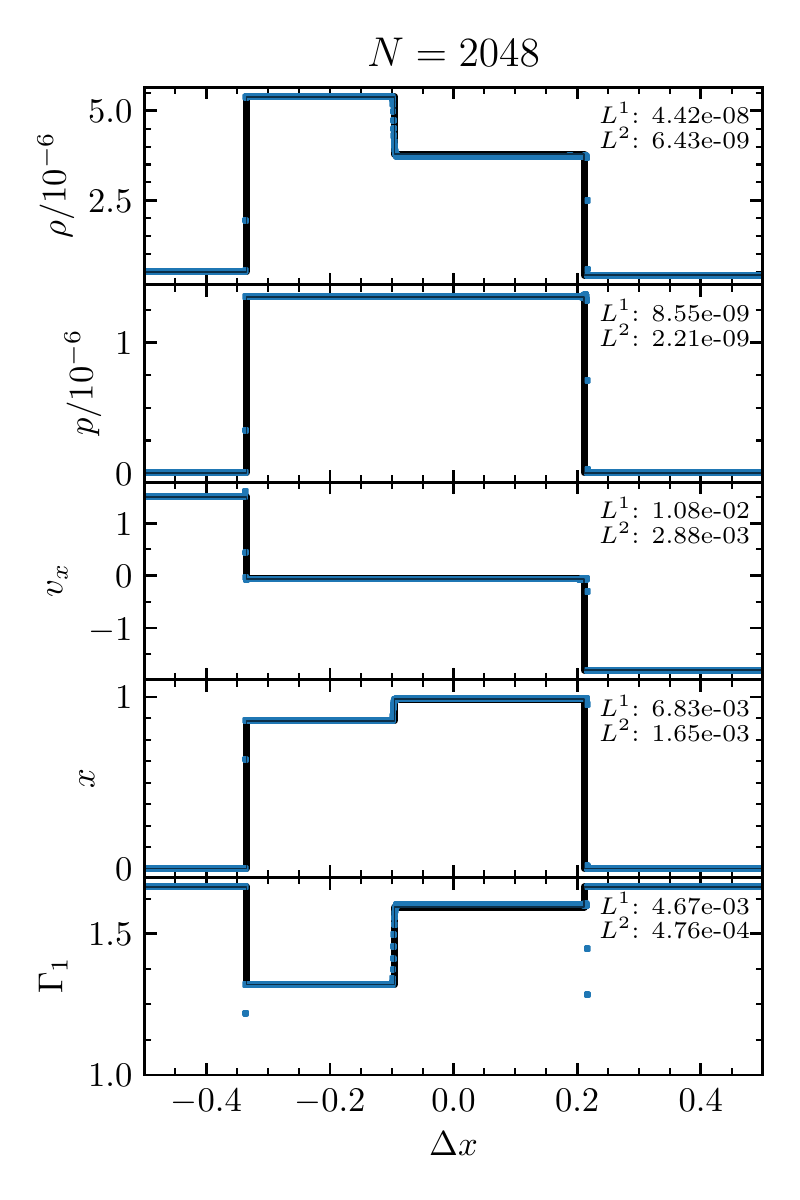}
	\end{center}
	\caption{Same as Fig.~\ref{fig:test01} but for Riemann test 4 (see Table~\ref{tab:tests}).
		\vspace{1in}
	}
	\label{fig:test04}
\end{figure*}

\begin{figure*}
	\begin{center}
		\includegraphics[width=.5\linewidth]{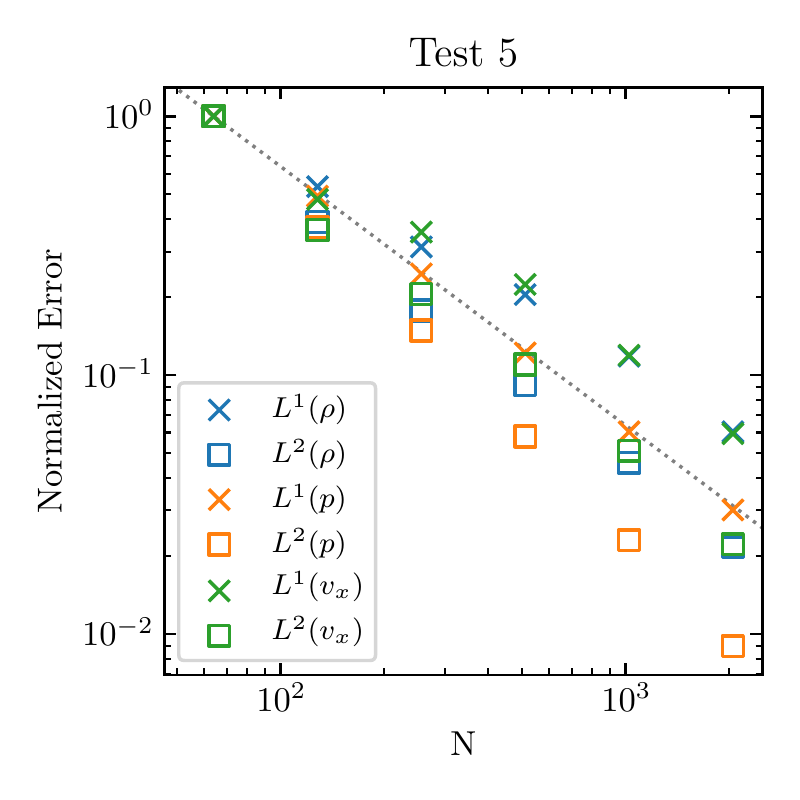}\\
		\includegraphics[width=.49\linewidth]{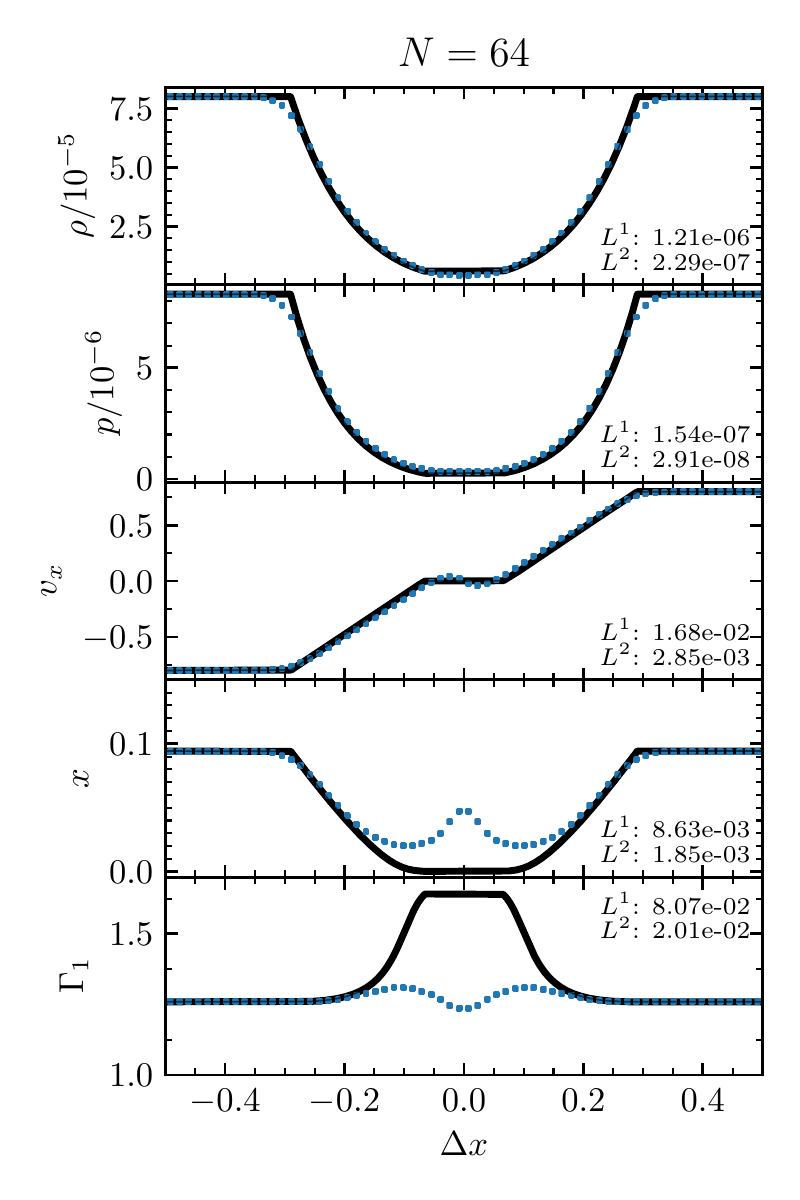}
		\includegraphics[width=.49\linewidth]{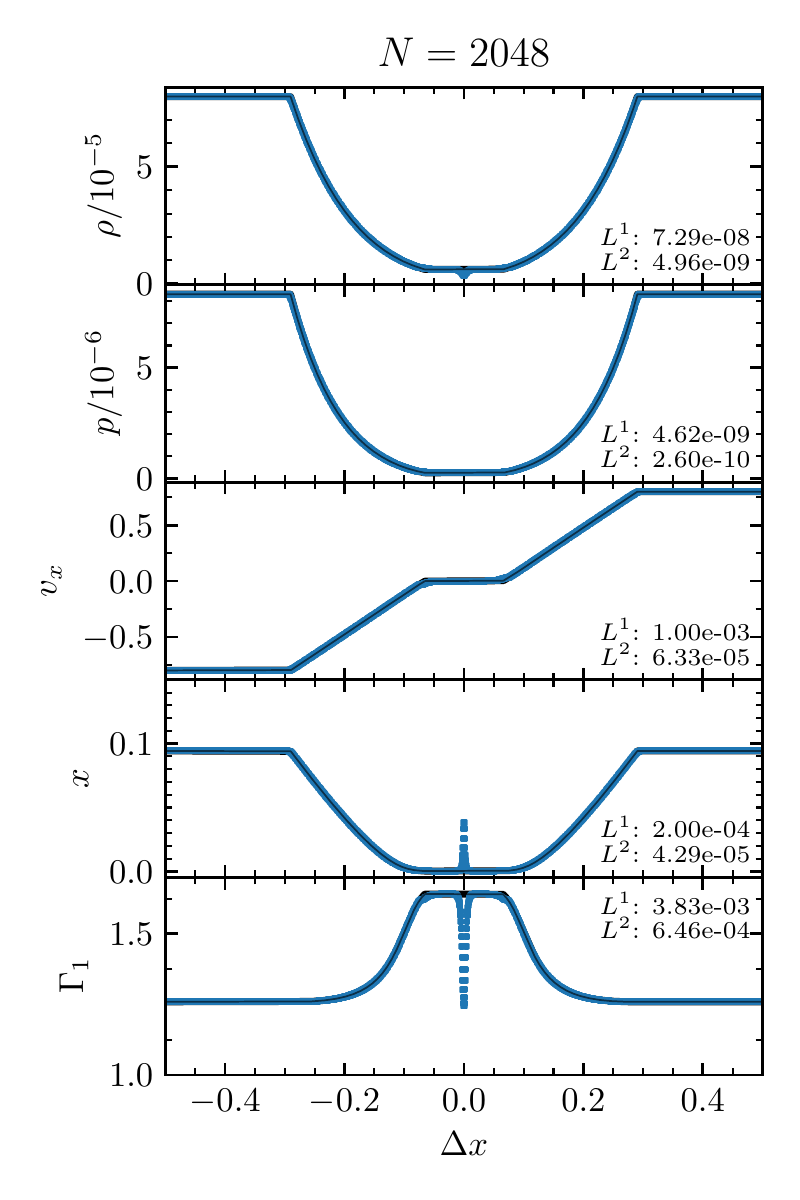}
	\end{center}
	\caption{Same as Fig.~\ref{fig:test01} but for Riemann test 5 (see Table~\ref{tab:tests}).
		\vspace{1in}
	}
	\label{fig:test05}
\end{figure*}

\begin{figure*}
	\begin{center}
		\includegraphics[width=.5\linewidth]{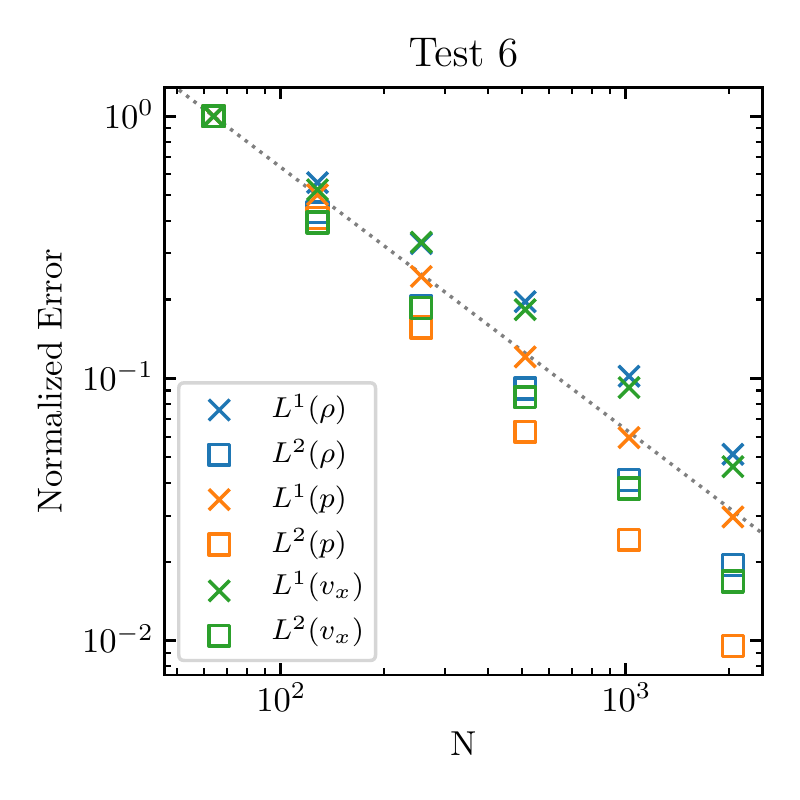}\\
		\includegraphics[width=.49\linewidth]{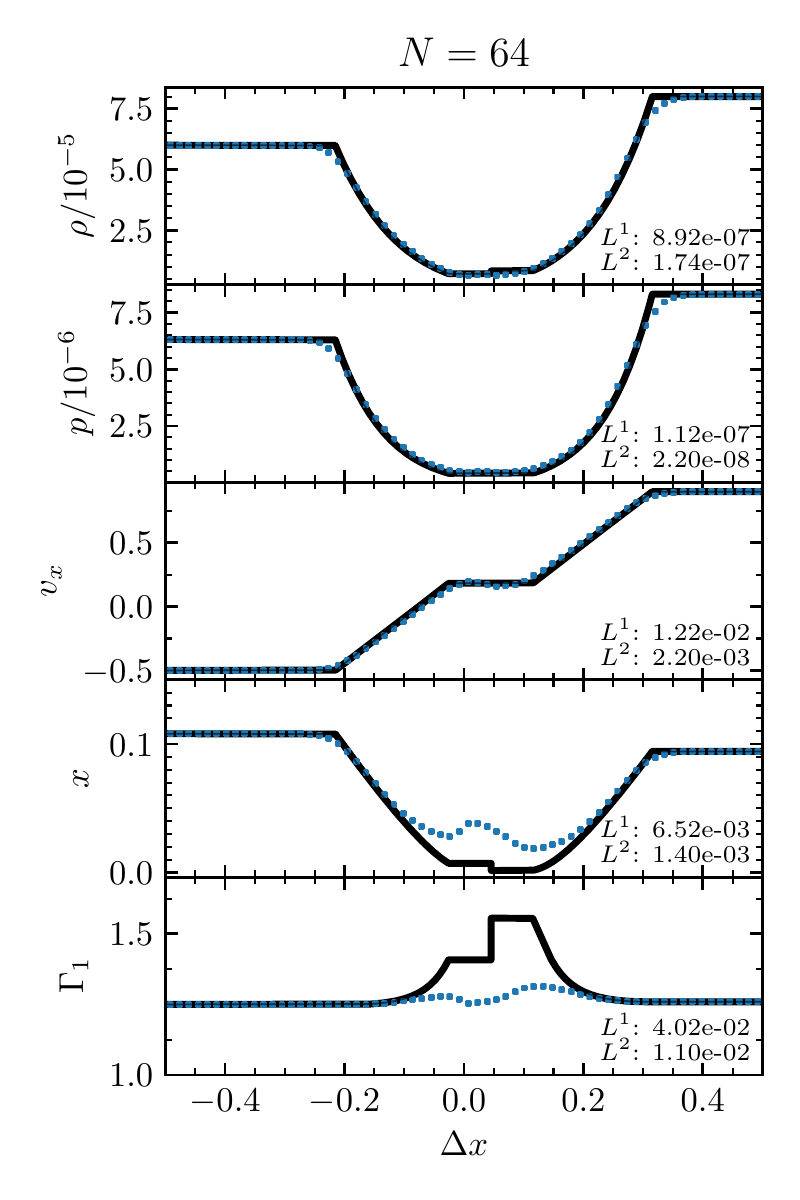}
		\includegraphics[width=.49\linewidth]{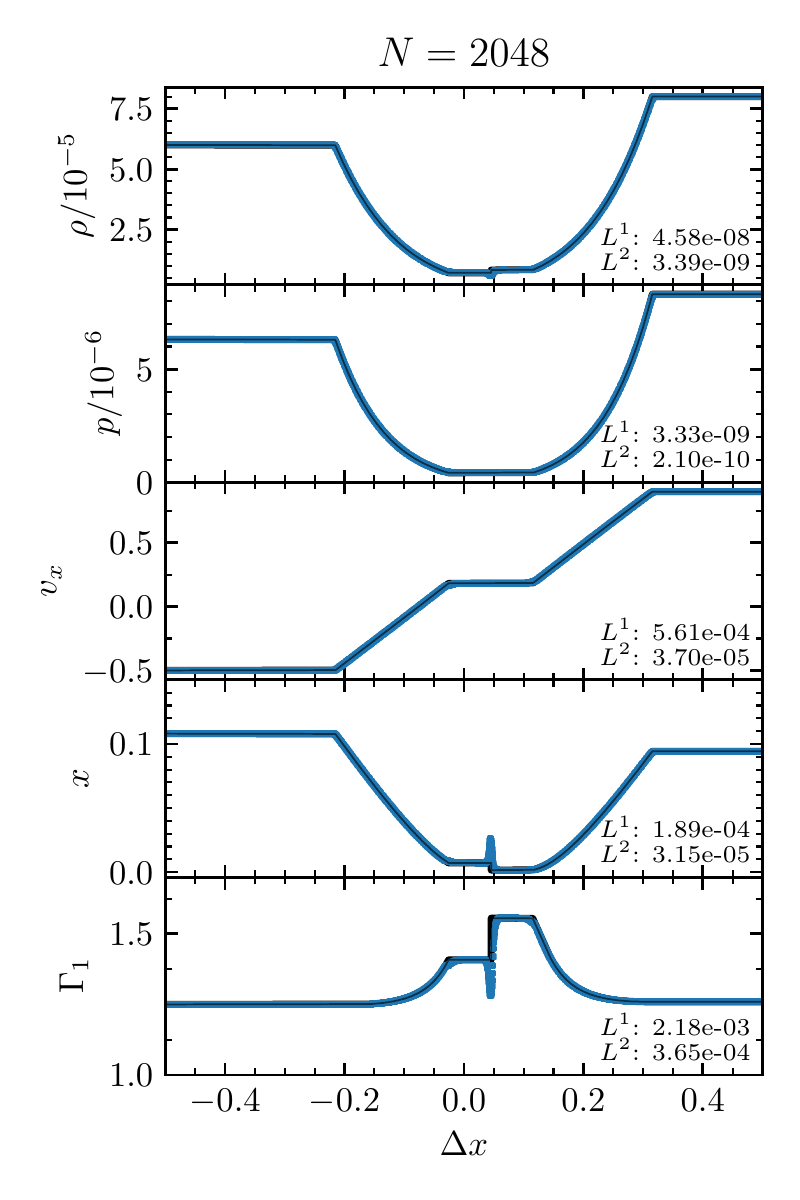}
	\end{center}
	\caption{Same as Fig.~\ref{fig:test01} but for Riemann test 6 (see Table~\ref{tab:tests}).
		\vspace{1in}
	}
	\label{fig:test06}
\end{figure*}

\begin{figure*}
	\includegraphics[width=1\linewidth]{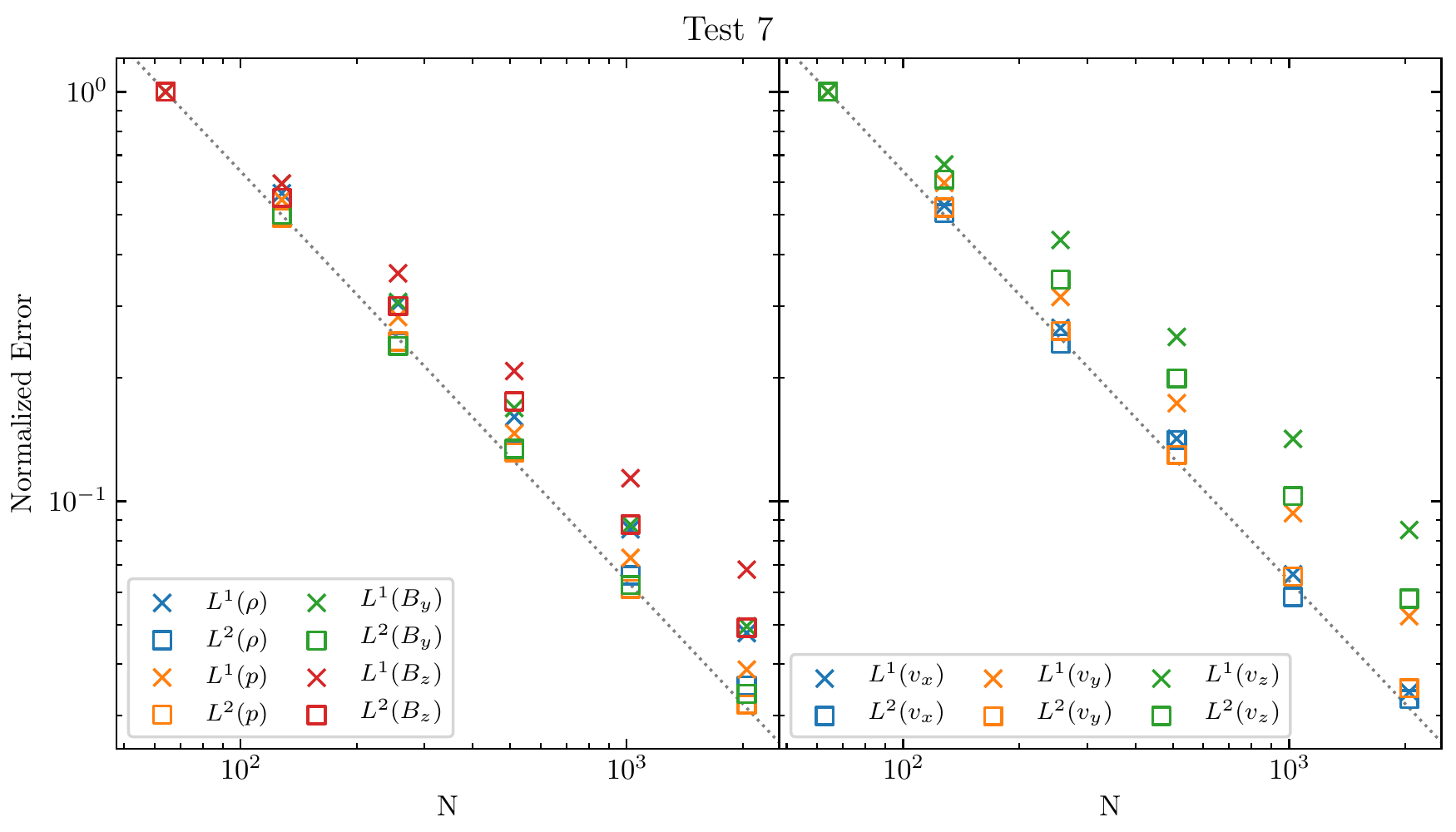}
	\vspace*{-1.2em}
	\caption{$L^1,\,L^2$ errors for various fluid parameters for test 7 (see Tables~\ref{tab:mhd-ic} and \ref{tab:mhd_err}), as a function of number of cells ($N$) normalized by their value at $N=64$. The dotted gray line indicates a linear trend.
	}
	\label{fig:test07a}
\end{figure*}

\begin{figure*}
	\noindent \centering{\Large Test 7}\\
	\includegraphics[width=.49\linewidth]{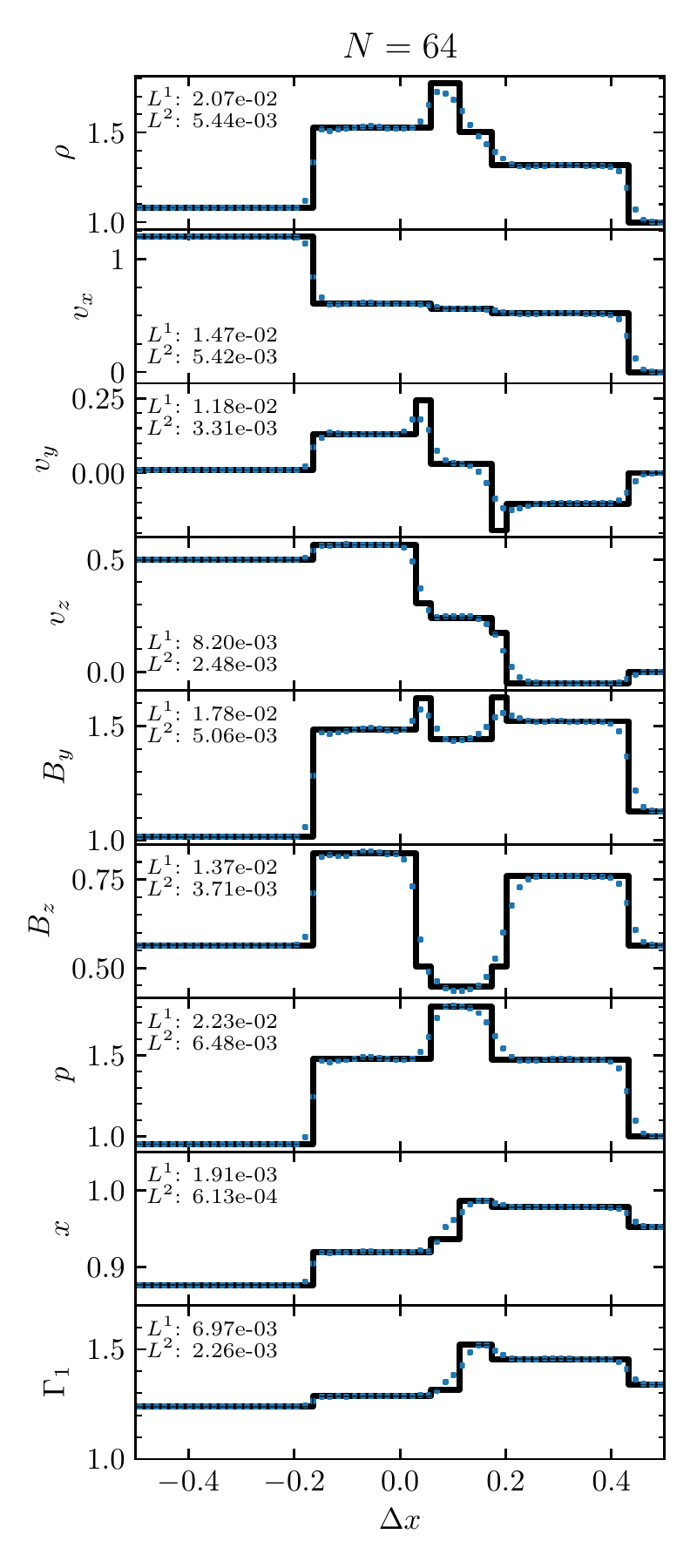}
	\hfill
	\includegraphics[width=.49\linewidth]{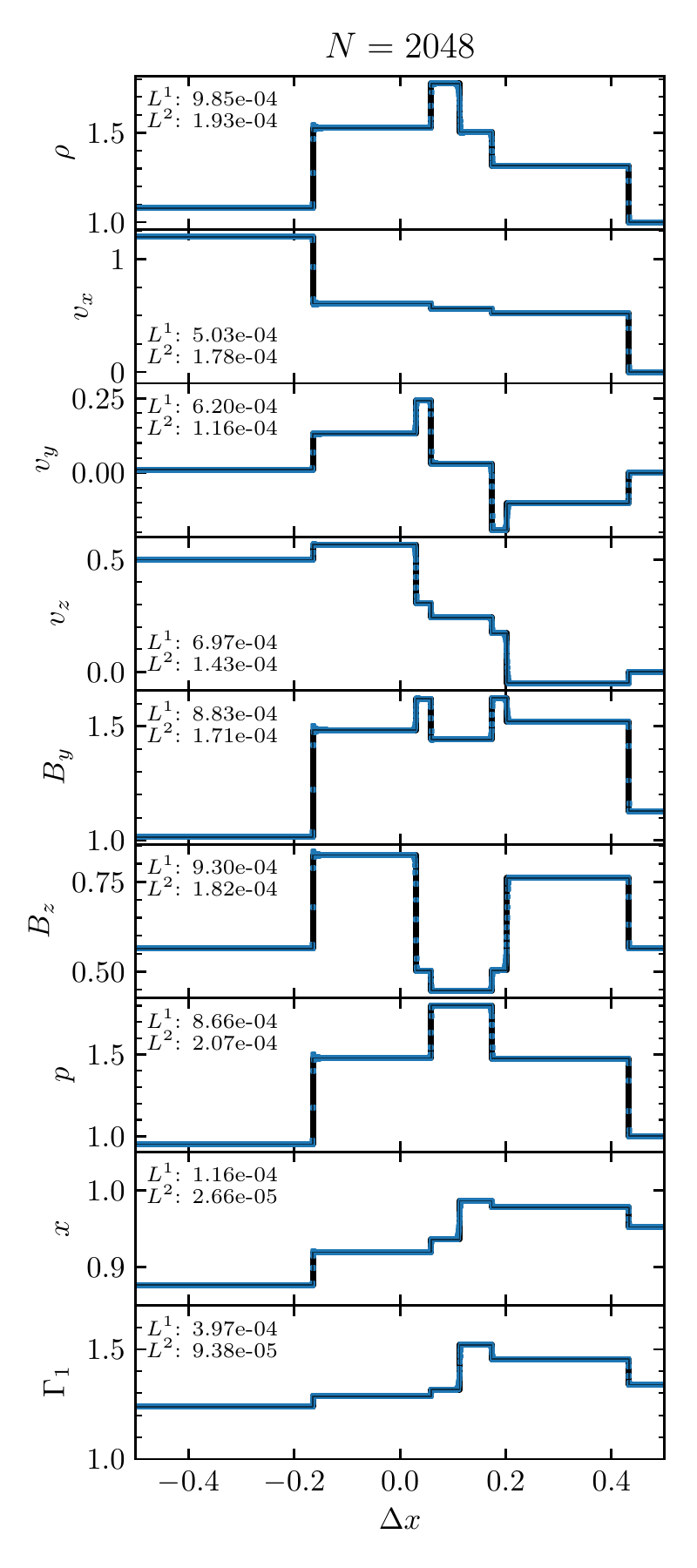}
	\vspace*{-1em}
	\caption{Profiles of fluid parameters for test 7 (see Table~\ref{tab:mhd-ic}) with the \athena results (blue points) and exact solution (black line) at $t=\Delta t$ for $N=64$ (left) and $N=2048$ (right).
	}
	\label{fig:test07b}
\end{figure*}

\begin{figure*}
	\includegraphics[width=.49\linewidth]{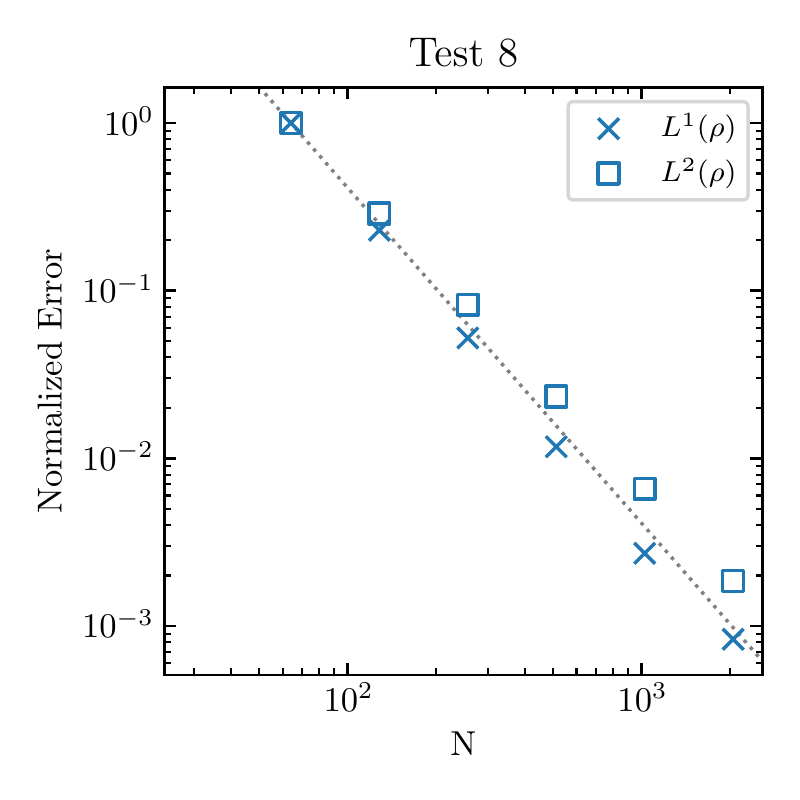}
	\hfill
	\includegraphics[width=.49\linewidth]{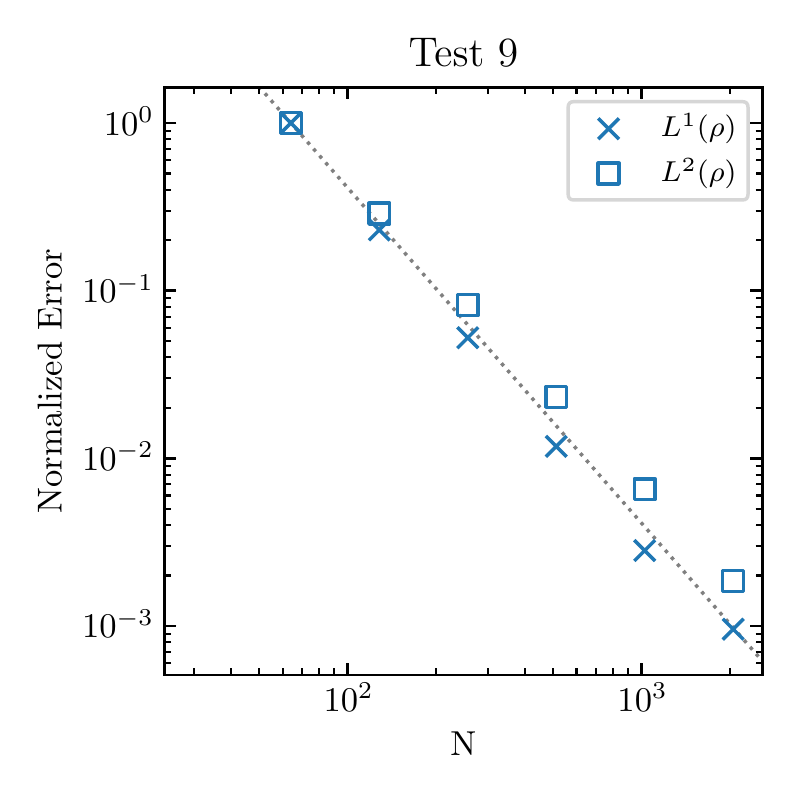}
	\vspace*{-1.2em}
	\caption{$L^1,\,L^2$ errors for density for the linear wave tests 8 and 9 (see Section~\ref{sec:lin-wave_test}), as a function of number of cells ($N$) normalized by their value at $N=64$. The dotted gray line indicates a quadratic trend. The normalized errors of other fluid quantities are not plotted because they are within $1\%$ of the normalized density errors.
	}
	\label{fig:tests8-9}
\end{figure*}

In addition to Riemann problem tests, we also ran two 1D periodic linear wave tests using our hydrogen EOS: one HD and one MHD test.
In these tests we initialize  a wave in an eigenmode of the (M)HD equations and run the simulation for exactly one wave crossing time. We then compare the final solution to the initial conditions.
For both of these tests we set the background density ($\rho_0$), temperature ($T_0$) and velocity ($\mathbf{v}_0$) to
\begin{align}
\rho_0&=10^{-7} \rho_{\rm u}\\
T_0&=0.0435 \, T_{\rm u}\\
\mathbf{v}_0&=\left\lbrace 0,0,0 \right\rbrace;
\end{align}
all other background quantities (e.g. $p_0$) are computed from these. This choice of $T_0$ is within $1\%$ of the value of $T$ that maximizes $\left|\partial\Gamma_1/\partial T\right|$ at $\rho=\rho_0$. We initialized the fluid quantities $f\in\left\lbrace \rho, \rho v_x, \rho v_y, \rho v_z, E, B_y, B_z\right\rbrace$ with a sine wave:
\begin{align}
f\left(x\right)&=f_0+\delta f \sin\left(\dfrac{2\pi x}{\Delta x}\right),
\end{align}
where $\delta f$ is the amplitude of the wave, and $\Delta x$ is the size of the simulation domain. We set
\begin{align}
\delta\rho=10^{-6} \rho_0.
\end{align}
For the HD test the other $\delta f$ quantities are chosen such that the wave is a left-moving ($-\hat{x}$) sound wave, with the relative amplitudes given by the coefficients of the right eigenvector corresponding to the eigenvalue of $-a$ \citepalias[see e.g. Appendix B of][]{}.
Similarly for the MHD test, the wave is a left-moving fast magnetosonic wave \citep[see][for the MHD eigenvectors]{doi:10.1063/1.4851415}, with a background magnetic field ($\mathbf{B}_0$) of
\begin{align}
\mathbf{B}_0=\sqrt{p_0}\times\left\lbrace 1,\sqrt{2},1/2 \right\rbrace,
\end{align}
with $B_x$ constant, i.e. $B_x(x)=B_{x,0}$.
As the linear waves do not contain discontinuities, the expected convergence is quadratic i.e.  $L^1\left(f\right)\propto1/N^2$, and $L^2\left(f\right)\propto1/N^2$, where N is the number of cells. To test convergence, these tests are run with $N=64, 128, 256, 512, 1024, 2048$ cells. As before, we set the CFL number to $0.4$.

\subsection{Helmholtz HD Tests}
\label{sec:helm_tests}

A commonly used non-ideal EOS used in astrophysical fluid dynamics (particularly in stellar interiors and supernovae) is the Helmholtz EOS \citep{helmeos}, based on the Helmholtz free energy. 
Despite its wide usage, there exists only one previous work \citep{2015ApJS..216...31Z} with HD tests showing convergence to a known solution. 
We add to this work by computing exact solutions to Riemann problems with the Helmholtz EOS in different parameter space.
We then conduct a proper HD convergence test using this EOS by implementing it in \athena and comparing the simulation results to the exact solutions. 
Even though this EOS has a tabular component to it, we do not utilize the tabular EOS formalism in Section~\ref{sec:eos-table}.
Instead, we implemented this EOS as a function of density and temperature exactly as described by \citet{helmeos} and preform a root-find\footnote{We use a combination of bisection, secant, and Newton–Raphson methods.} to determine the temperature at each EOS call. The last recovered temperature is used as an initial guess for the next EOS call. The inclusion of the Helmholtz EOS also demonstrates the flexibility and extensibility of our EOS framework.

\begin{figure*}
	\begin{center}
		\includegraphics[width=.5\linewidth]{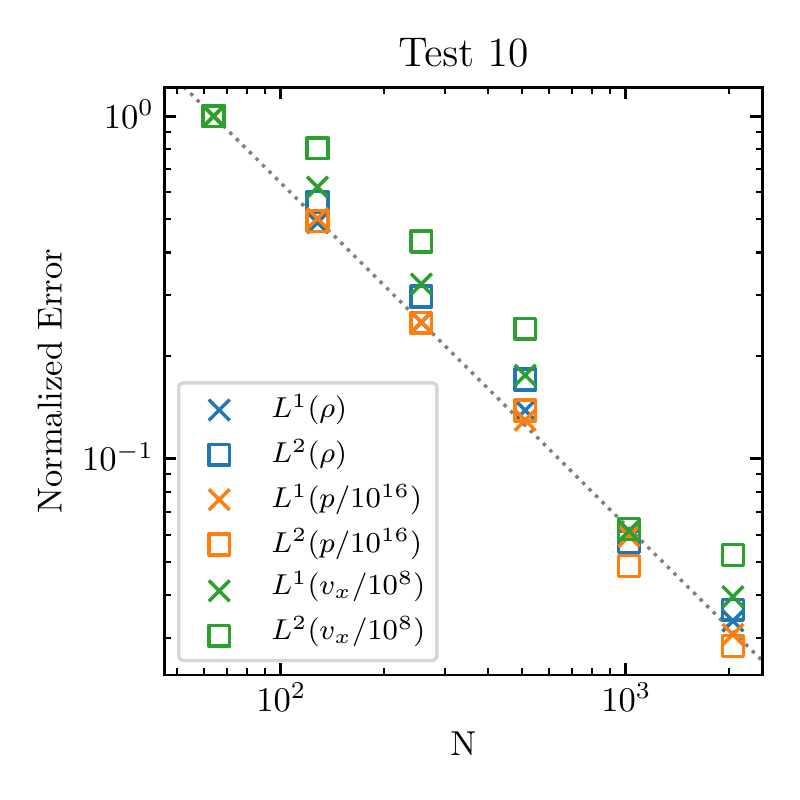}\\
		\includegraphics[width=.49\linewidth]{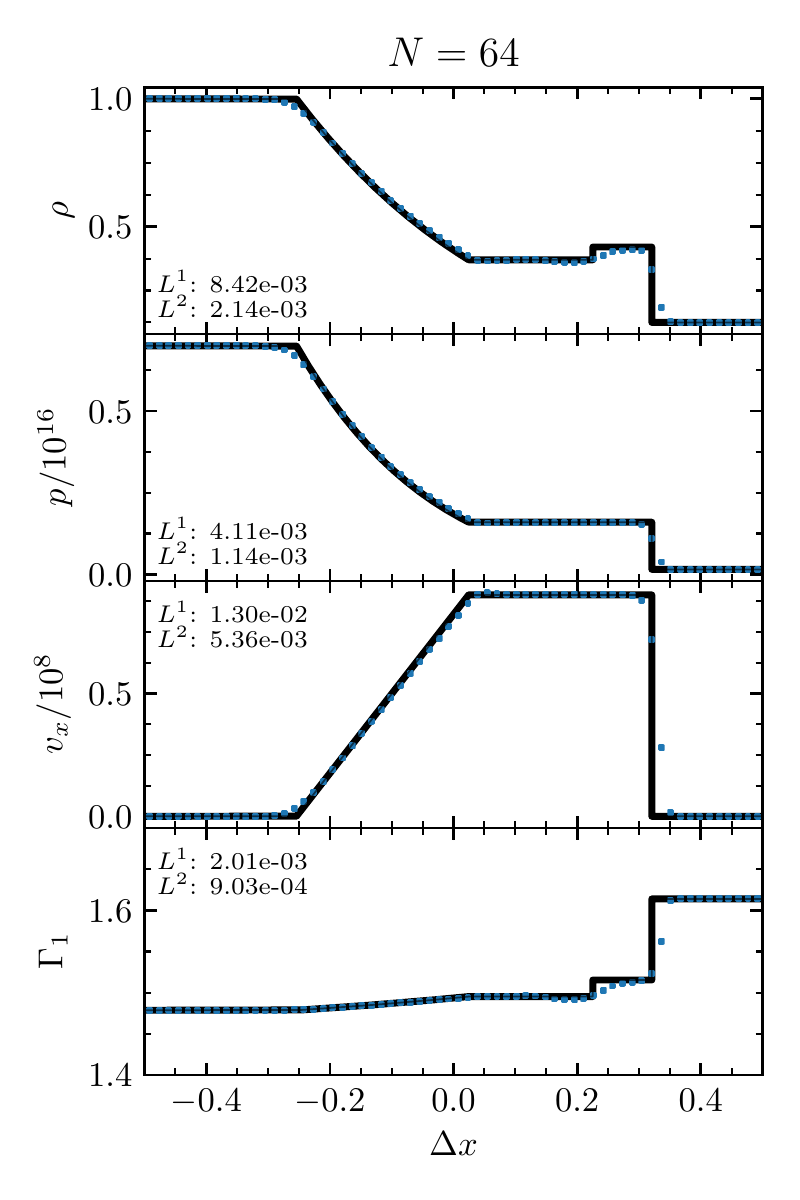}
		\includegraphics[width=.49\linewidth]{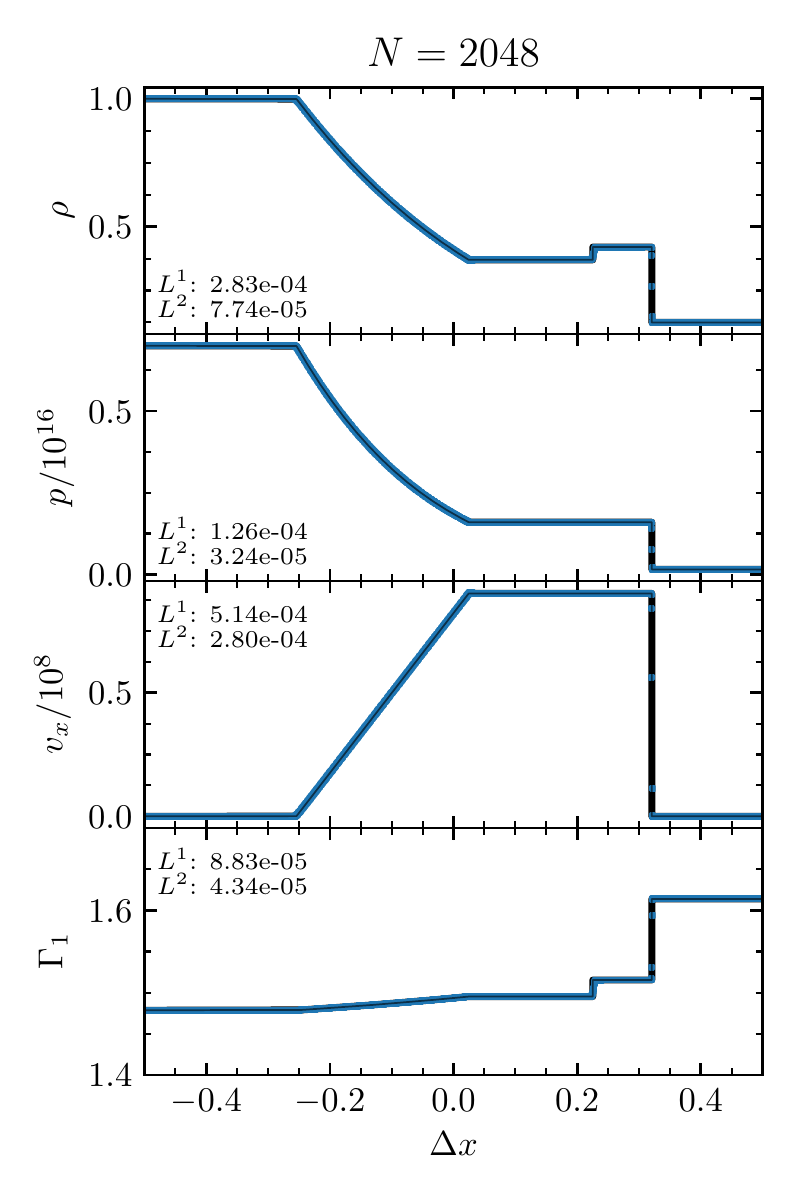}
	\end{center}
	\caption{Same as Fig.~\ref{fig:test01} but for Riemann test 10 (see Table~\ref{tab:helm_tests}).
		\vspace{1in}
	}
	\label{fig:test10}
\end{figure*}

\begin{figure*}
	\begin{center}
		\includegraphics[width=.5\linewidth]{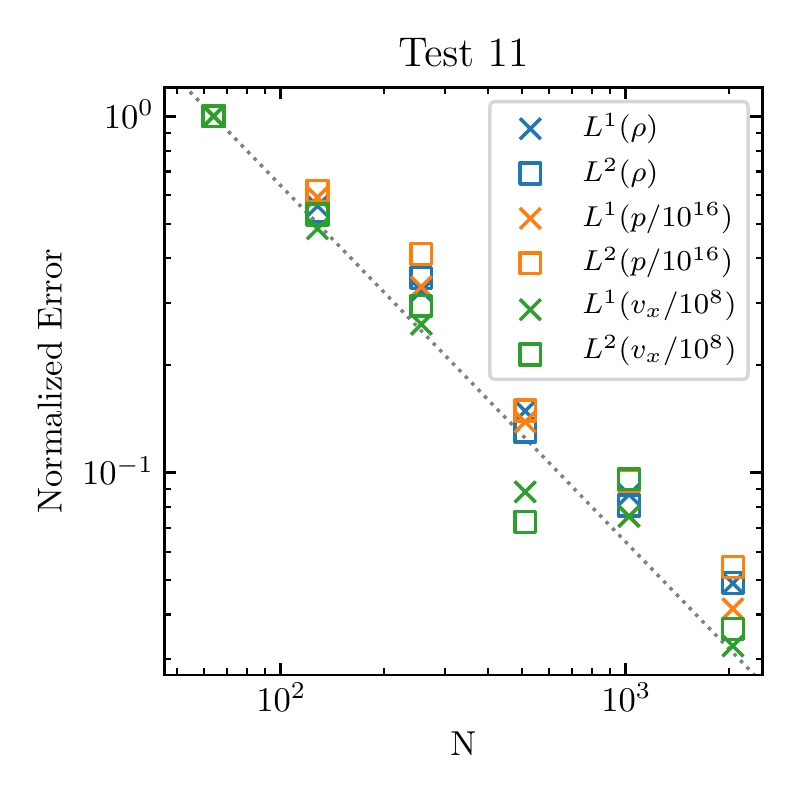}\\
		\includegraphics[width=.49\linewidth]{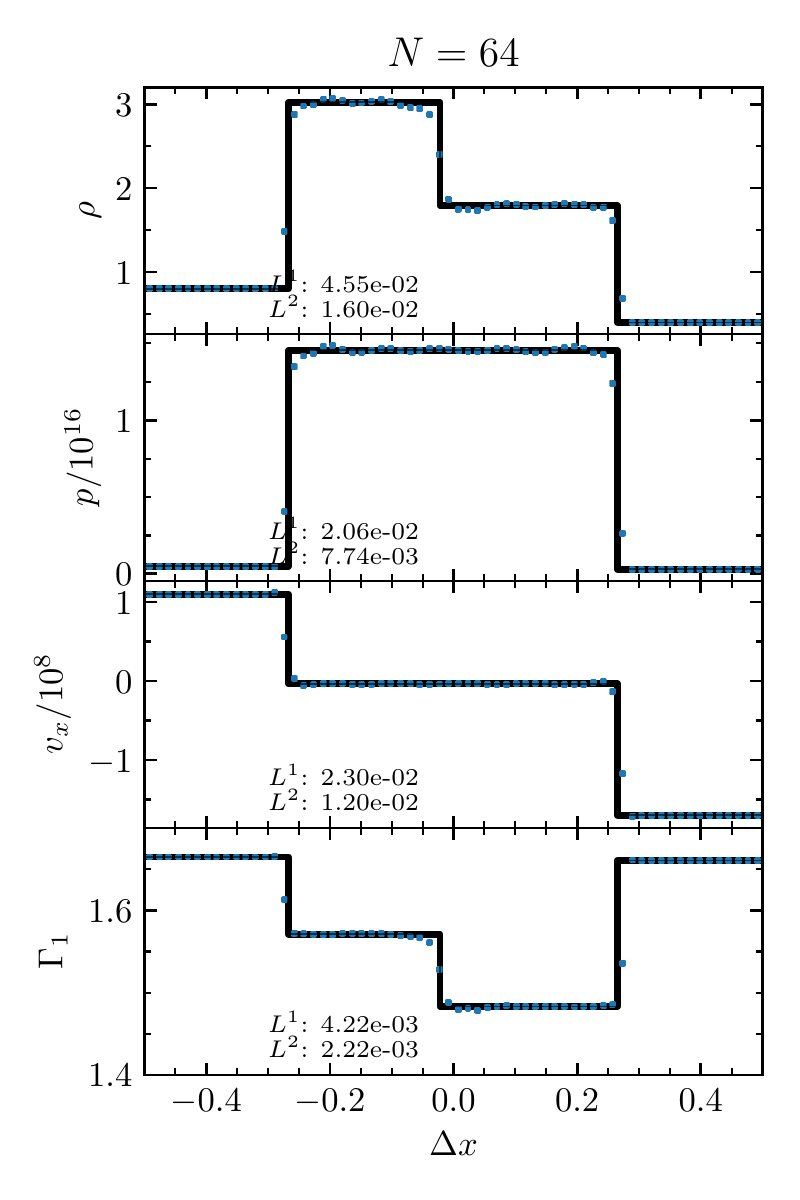}
		\includegraphics[width=.49\linewidth]{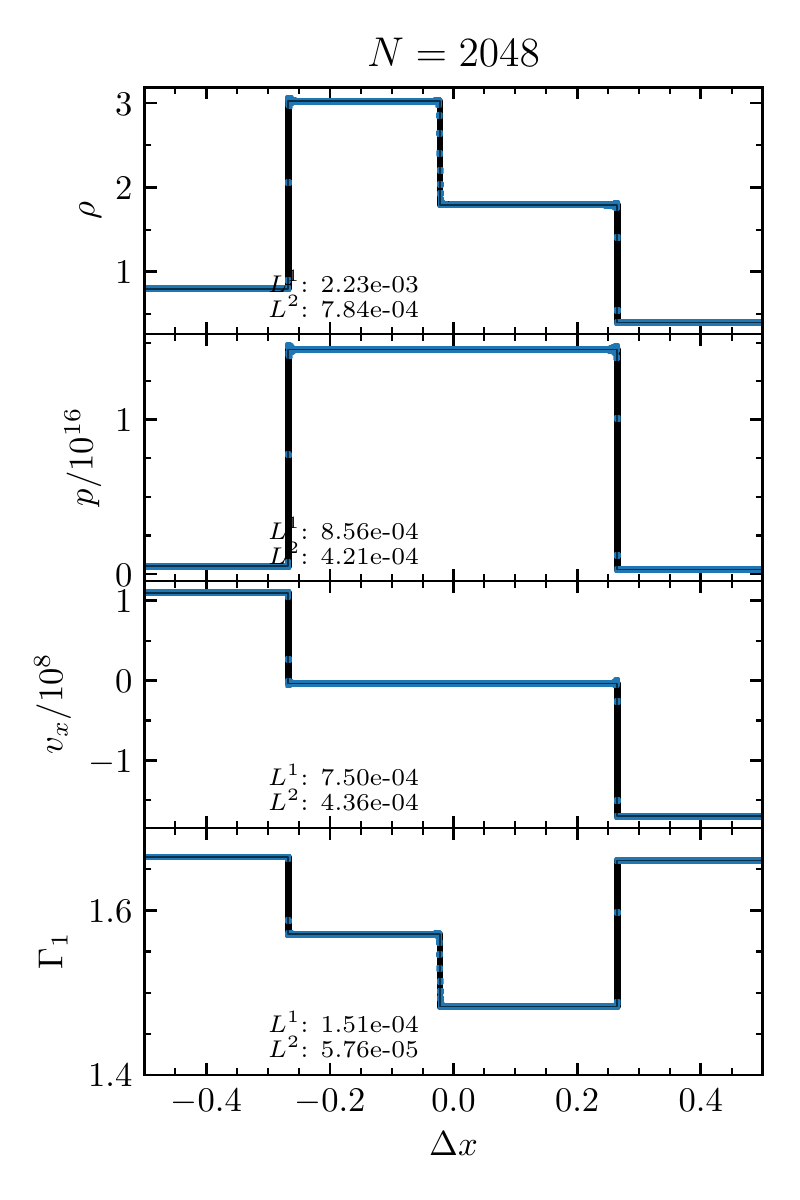}
	\end{center}
	\caption{Same as Fig.~\ref{fig:test01} but for Riemann test 11 (see Table~\ref{tab:helm_tests}).
		\vspace{1in}
	}
	\label{fig:test11}
\end{figure*}

In developing a Riemann problem test we restrict ourselves to (density-temperature) regimes where the sound speed is non-relativistic and the pressure is not dominated by degeneracy pressure. 
The EOS framework described here is not capable of relativistic calculations, giving rise to our first constraint,
while the second constraint is more numeric in origin, as it is problematic to invert $p=p(\rho,T)$ for temperature when degeneracy pressure dominates. As in our previous tests we also wish to generate tests with a noticeable variation of $\Gamma_1$. This set of constraints precludes double rarefaction-wave tests because $\Gamma_1$ varies negligibly along adiabats (which rarefaction-waves follow) in the non-degenerate non-relativistic regime. We are left with a Sod-like and a double shock test, tests 8 and 9 respectively (see Table~\ref{tab:helm_tests} and Figs.~\ref{fig:test10} and \ref{fig:test11}). As before we set the CFL number to $0.4$. We limit ourselves to two Helmholtz EOS tests, as the variation of $\Gamma_1$ is small compared to the hydrogen EOS tests, making Helmholtz less challenging.

\section{Results and Discussion}
\label{sec:dicussion}

\subsection{Hydrogen HD Riemann Tests}

With the exception of the double shock tests (3 and 4), all tests show convergences rates close to linear, i.e. the best-case rate.  We elaborate on the outcomes and details of these tests below.

\subsubsection{Sod-Like}

Tests 1 and 2 show convergence very close to linear (see Figs.~\ref{fig:test01} and \ref{fig:test02}). 
\mc{
Previous work \citep[e.g.][]{Banks_Aslam_Rider_2008} has shown that the high resolution convergence for contact discontinuities is sub-linear, implying that these errors are sub-dominant here.
}
The two discontinuities (contact in the middle and shock to the right) are spread out over several grid cells (an effect typically called numerical viscosity), but track the actual discontinuities well at high resolution. Both of these tests show overshooting of $v_x$ at the transition from the rarefaction wave to the $L*$ state (recall Fig.~\ref{fig:fan}), which manifests as a noticeable-but-small ``notch" feature at $N=2048$. Similarly, test 1 shows undershooting of $\rho$ and $p$ at the same transition and similar ``notch" features.

\subsubsection{Shock-Shock}

Tests 3 and 4 exhibit the worst convergence (only at high resolution) compared to the other tests by a significant margin. Both of these tests also show worsening convergence at higher resolution ($N\gtrsim256$), with errors for test 3 nearly constant for $512\leq N \leq 2048$. Low amplitude oscillations of $\rho$ and $p$ about the exact solution are present (most noticeable at $N=64$). At $N=2048$ these oscillations are reminiscent of ringing artifacts and the Gibbs phenomenon especially at the location of the rightmost shock. \citetalias{Chen19} have noted similar issues, and we have seen comparable results in analogous ideal gas EOS tests, in terms of convergence and deviations from the solution at a fixed resolution. 

This lack of convergence may by indicative of the limitations of a finite volume methods without front tracking; specifically, when a cell straddles a discontinuity, utilizing the volume averaged state likely results in errors that cannot be diminished with increasing resolution. This is especially noticeable at the forward and reverse shocks in tests 3 and 4 (see Figs.~\ref{fig:test03} and \ref{fig:test04}), where points are seen well below the correct $\Gamma_1$ curve. This is because $\Gamma_1$ is non-monotonic here, causing the cell-averaged density and pressure (through the EOS) to give a $\Gamma_1$ lower than the $\Gamma_1$ on either side of the shock. This under-prediction of $\Gamma_1$ is roughly the same for $N=64$ and $N=2048$, suggesting that this is related to the lack of convergence at high resolution. \mc{We also note that there is visible diffusion of the contact discontinuity, which could suggest that these tests are being affected by the sub-linear convergence expected of contact discontinuities \citep[see e.g.][]{Banks_Aslam_Rider_2008}.}

Despite these issues, even the $N=64$ case (for both tests) well approximates the exact solution, indicating that our code preforms reasonably well. Additionally, these test were designed as a worse case scenario, so the fact that at $N=64$ the recovered solution is close to the exact solution is reassuring.

As alluded to in Section~\ref{sec:Riemann}, these tests seem to be sensitive to the choice of wave-speed estimators. For the HLLC Riemann solver we found that the PVRS estimator performed significantly better then the Roe averaged method (see Appendix~\ref{sec:wavespeeds} for details on these wave-speed estimators). We hope that these tests could be utilized for further examination of the impact of different wave-speed estimators.

\subsubsection{Double-Rarefaction}

Tests 5 and 6 both show close to linear convergence, and small errors (see Figs.~\ref{fig:test05} and \ref{fig:test06}). A small deviation in density is noticeable in both tests near the middle of the simulation. This central spike is a common feature in numerical solutions of double-rarefaction wave tests \citep[see e.g.][]{ToroBook, Chen19}. It is particularly surprising how significantly affected the ionization fraction ($x$) and $\Gamma_1$ are effected here (especially at $N=64$). This is because the test is tuned to be at the ionization edge, making these parameters very sensitive to pressure and density. Despite the large deviations of $x$, and $\Gamma_1$, the remainder of the plotted fluid parameters are well behaved and remain close to the exact solution.

\subsubsection{HLLE}

We experimented with the HLLE\footnote{Also called HLL.} Riemann solver using a variety of wavespeed estimators. None of the wavespeed estimators we tried effected the outcomes of tests 1, 2, 5 and 6 (this is also true for HLLC). For these tests, HLLE had found $L^1$ and $L^2$ errors a few tens of per cent worse than HLLC. For tests 3 and 4 HLLE did not converge to the correct solution for any of the wavespeed estimators we tried\footnote{However, HLLE with the Roe averaged wavespeed estimator (see Section~\ref{sec:roe}) does converge to the correct solutions on the ideal gas EOS analogs of tests 3 and 4.}; although there may exist a wavespeed estimator which does enable HLLE to properly converge.

\subsection{Hydrogen MHD Riemann Test}
Test 7 shows near linear convergence for all the fluid variables and the recovered solution well approximates the exact solution. We also ran tests $1-6$ where we explicitly use the MHD equations with $\mathbf{B}=0$ and the HLLD Riemann solver. With the exception of tests 3 and 4, all results and errors were consistent to the results for HLLC (using the Euler equations). Tests 3 and 4 had slightly worse convergence, and the deviations (i.e. $L^1$ and $L^2$) of the recovered solution compared to the exact one were worse by a factor of up to three. This is due to the different wave-speed estimator; when we change HLLC to use the estimator that is currently used by HLLD, similar results are recovered. Unfortunately, the PVRS wave-speed estimator (see Appendix~\ref{sec:wavespeeds}) is not readily extendable to MHD. This indicates that further work is needed on testing and deriving different wave-speed estimators for HLLD.

\subsection{Hydrogen Linear Wave Tests}

The linear wave tests (8 and 9) demonstrate the ability of our modifications of \athena to converge on continuous (M)HD solutions. Here we see near quadratic convergence (see Fig~\ref{fig:tests8-9}), as expected for (M)HD problems without discontinuities. This indicates that the new EOS capabilities of \athena can accurately evolve (M)HD and properly converge on the correct solution for problems without discontinuities.

\subsection{Helmholtz HD Tests}

Both the Sod-like and double shock tests (tests 8 and 9 respectively) show good convergence and accurately reproduce the exact solution. At $N=2048$ test 10 shows no significant deviation from the exact solution, and even the $N=64$ case tracks the exact solution rather well. 
Test 11 exhibits similar features compared to the analogues hydrogen tests (3 and 4). At $N=64$ oscillations about the exact solution are present, albeit with small amplitudes. The ringing artifacts are again present at $N=2048$ around the two shock discontinuities. 

We note that test 11 has more accurate results and better convergence compared to the analogues hydrogen tests, indicating that the hydrogen EOS provides more strenuous tests of our modifications to \athena. This is likely because the hydrogen EOS has more rapid and substantial variations in $\Gamma_1$, and discontinuities can span regions where $\Gamma_1$ is non-monotonic. Our tests using the Helmholtz EOS show that this EOS is well behaved for (M)HD applications with Godunov-type codes, and that our new framework is capable of handling a variety of EOS.

\section{Summary and Conclusion}
\label{sec:conclusions}

In this paper, we described new modifications to the \athena framework which enables the use of general EOS in solving non-relativistic (M)HD problems. 
This required modifications to EOS calls and the (HLLC, and HLLD) Riemann solvers within \athena, and we were unable to find a modification to HLLE that enabled it to properly converge to the solutions of tests 3 and 4.

Due to the lack of previously existing tests, we generated a series of (M)HD tests utilizing a non-trivial EOS based on a hydrogen gas, and two HD tests using the Helmholtz EOS \citep{helmeos}. We then use these tests to verify the accuracy and convergence rate of our modifications to \athena. For the most part our code recovered solutions close to the exact ones and converged as expected. This was less true for the hydrogen EOS double-shock tests (3 and 4), but the deviations from the exact solution were still small in these cases. These double-shock tests may also demonstrate a limit of the finite volume method: near the forward and revere shock-discontinuities passing the cell-averaged fluid state to the EOS results in $\Gamma_1$ values that are significantly lower compared to $\Gamma_1$ on either side of te shock.

In our investigation we discovered that wave-speed estimator used within the Riemann solver can make a substantial difference in both the convergence rate and overall accuracy, in particular for tests 3 and 4. Accordingly, we suggest further investigation into wave-speed estimators, especially for HLLD.

By making our newly developed tests freely available to other code developers and by including our EOS framework into the publicly available version of \athena, we hope to enable a wide variety of research involving astrophysical fluids where the complexities of a realistic EOS have a substantial impact. 

\acknowledgments
\section*{Acknowledgments}

We thank James Stone, Kyle Felker, Kengo Tomida, Yan-Fei Jiang, Omer Blaes, Chris White, Zhuo Chen, 
and Goni Halevi for their useful discussions and insight generated from their work.
We also thank the anonymous
referee who’s feedback lead to improvements of this paper.
MC gratefully acknowledges support from the Institute for Advanced
Study, NSF via grant AST-1515763, and NASA via grant
14-ATP14-0059.

\software{
	\athena \citep{athena},
	\texttt{Matplotlib} \citep{Matplotlib},
	\texttt{NumPy} \citep{Numpy}
}

\bibliographystyle{aasjournal}
\bibliography{citations}

\appendix

\section{Wave-speed Estimates}
\label{sec:wavespeeds}

\subsection{Roe Average}
\label{sec:roe}

Before the work presented here, \athena used the Roe average to approximate the middle state as a means to estimate the extremal wave-speeds \citep[see e.g. Sections 10.5.1 and 11.3.3 of][]{ToroBook}.
The Roe average middle state gives
\begin{align}
\tilde{\rho}&=\sqrt{\rho_{L}\rho_{R}}\\
\tilde{v}_x&=\frac{\sqrt{\rho_{L}} v_{x,L}+\sqrt{\rho_{R}} v_{x,R}}{\sqrt{\rho_{L}}+\sqrt{\rho_{R}}}\\
\tilde{H}&=\frac{\sqrt{\rho_{L}} H_{L}+\sqrt{\rho_{R}} H_{R}}{\sqrt{\rho_{L}}+\sqrt{\rho_{R}}}\\
\label{eqn:RiemannAsq}
\tilde{a}^2&=a^2\left(\tilde{\rho}, \tilde{h}\right),
\end{align}
where $H=p/\rho+e+v^2/2$ is the total specific enthalpy, $h=H-v^2/2$ is the \emph{internal} specific enthalpy, and the Tilda (e.g $\tilde{H}$) denotes the Roe average of the quantity underneath.
Note, that this requires a fourth EOS function, $a^2\left(\rho, h\right)$, in addition to the three previously specified by Eqns.~\ref{eqn:eos1}-\ref{eqn:eos3}. 
For an ideal gas, Eqn.~\ref{eqn:RiemannAsq} simplifies to
\begin{equation}
\tilde{a}^2=(\gamma-1)\tilde{h}.
\end{equation}
Using these results, the approximations of the extremal wave-speeds are
\begin{align}
\lambda_{\rm min}&=\text{min}\left(v_{x,L}-a_L,\tilde{v}_x-\tilde{a}\right)\\
\lambda_{\rm max}&=\text{max}\left(v_{x,R}+a_R,\tilde{v}_x+\tilde{a}\right).
\end{align}

Another generalization of the Roe average wave-speed estimator is given by \citet{HU09}, which utilizes a different way of computing $\tilde{a}^2$ (Eqn.~\ref{eqn:RiemannAsq}). In this formalism four EOS functions are still needed. This method gives comparable results to our above extension of the Roe average wave-speed estimator, but worse errors and convergence compared to the below PVRS Method for tests 3 and 4.

\subsection{PVRS Method}
After experimenting with a few wave-speed estimators we discovered that the primitive variable Riemann Solver (PVRS) method for estimating the wave-speed, described in Sections 9.3 and 10.5.2 of \citet{ToroBook}, resulted in significant reduction of errors and improved convergence in tests 3 and 4 (see Section~\ref{sec:tests} and Table~\ref{tab:tests}). Accordingly, we now use this method to estimate the extremal wave-speeds, as we describe here. 
The middle sates are estimated by
\begin{align}
\overline{\rho}&=\frac{1}{2}\left(\rho_{L}+\rho_{R}\right)\\
\overline{a}&=\frac{1}{2}\left(a_{L}+a_{R}\right)\\
p_{*} &=\frac{1}{2}\left(p_{L}+p_{R}\right)-\frac{1}{2}\left(v_{x,R}-v_{x,L}\right) \overline{\rho}\, \overline{a}\\
v_{x*} &=\frac{1}{2}\left(v_{x,L}+v_{x,R}\right)-\frac{1}{2}\left.\left(p_{x,R}-p_{x,L}\right) \middle/ \left(\overline{\rho}\, \overline{a}\right)\right.\\
\label{eqn:rho-mid-l}
\rho_{* L}&=\rho_{L} +\left(v_{x,L}-v_{x*}\right)(\overline{\rho} / \overline{a})\\
\label{eqn:rho-mid-r}
\rho_{* R}&=\rho_{R} +\left(v_{x*}-v_{x,R}\right)(\overline{\rho} / \overline{a})\\
\label{eqn:gamma-mid}
\Gamma_{*K}&=\dfrac{\rho_{* K}}{p_{*}}a^2\left(\rho_{* K}, p_{*}\right),
\end{align}
where $K$ is either $L$ or $R$.
We note that the last of these equations become trivial for an ideal EOS ($\Gamma_{*K}=\gamma$), making the computation of $\rho_{*K}$ unnecessary (Eqns. \ref{eqn:rho-mid-l} and \ref{eqn:rho-mid-r}).

With these computed we estimate the extremal wave-speeds as
\begin{align}
q_{K}&=\left\{\begin{array}{cc}{1} & {\text { if } p_{*} \leq p_{K}} \\ {\sqrt{1+\frac{\Gamma_{*K}+1}{2 \Gamma_{*K}}\left(p_{*} / p_{K}-1\right)}} & {\text { if } p_{*}>p_{K}}\end{array}\right.\\
\lambda_{\rm min}&=v_{x,L}-a_{L} q_{L}\\
\lambda_{\rm max}&=v_{x,R}+a_{R} q_{R}.
\end{align}

The added benefit of using the PVRS method over the Roe average method, is that it reduces the number of required EOS functions from four to three.

\section{Simple Hydrogen EOS}
\label{sec:eos}
To test our Riemann solvers in the simplest possible EOS which contains an ionization transition, we consider a plasma with only three species: electrons ($e^-$), neutral hydrogen (H$^0$), and protons/ionized hydrogen (H$^+$). We shall assume local thermal equilibrium (LTE) and consider only one reaction:
\begin{align}
\label{eqn:Hion}
{\rm H}^0 \rightleftharpoons {\rm H}^++e^-
\end{align}

To derive this EOS we need to know the relevant partition functions. $Z_{i,r}$ is the partition function per volume for the $r^{\rm th}$
ionized state of species $i$ (henceforth partition function will be used to mean partition function per volume). The partition function can be broken down into parts
\begin{equation}
Z_{i,r}=Z_{i,r}^{\rm bound} \times Z_{i,r}^{\rm nuc} \times Z_{i,r}^{\rm elec} \times Z_{i,r}^{\rm tr} \times Z_{i,r}^{\rm exct}.
\end{equation}
These parts are: internal bound states $Z_{i,r}^{\rm bound}$, nuclear spin $Z_{i,r}^{\rm nuc}$, electron spin and angular momentum $Z_{i,r}^{\rm elec}$, translation $Z_{i,r}^{\rm tr}$, and excitation $Z_{i,r}^{\rm exct}$. To further simplify the problem we shall assume that
\begin{align}
Z_{i,r}^{\rm bound}=1\;\;\text{for all }i,r.
\end{align}
While this is not technically true for H$^0$, it is a relatively small effect on the EOS. Additionally, this EOS is meant as a simple proof of concept and not for high precision applications.

We now list the partition functions for all the species. The partition functions not explicitly described are unity. The translational partition function has the same form for all species
\begin{equation}
Z_{i,r}^{\rm tr}=\left(\dfrac{2\pi m_ikT}{h^2}\right)^{3/2}=n_{\rm q}\left(\dfrac{m_i}{m_e}\dfrac{T}{T_{\rm ion}}\right)^{3/2},
\end{equation}
where
\begin{align}
n_{\rm q}&\equiv\left(\dfrac{2 \pi  m_e k T_{\rm ion}}{h^2}\right)^{3/2}=1.514892\times 10^{23}\text{ cm}^{-3}\\
T_{\rm ion}&\equiv\dfrac{1}{k}\dfrac{\alpha^2m_ec^2}{2}=157\,888\text{ K}.
\end{align}
and is the only ``partition function'' which is actually a partition function per volume. Physically, $T_{\rm ion}$ corresponds to the ionization energy of hydrogen, and at this temperature $n_{\rm q}$ is roughly the number density where the quantum degeneracy pressure of elections becomes important.
Note that we neglect differences in mass due to ionization level (i.e. $m_{i,r}=m_i$ for all $r$).
The remaining non-trivial partition functions are 
\begin{align}
Z_e^{\rm elec}&=2,\\
Z_{{\rm H},0}^{\rm elec}&=2,\\
Z_{{\rm H},1}^{\rm exct}&=\exp\left(-T_{\rm ion}/T\right),\\
Z_{{\rm H},r}^{\rm nuc}&=2,
\end{align}
while all the unspecified partition functions are unity.

To compute the EOS we must solve the Saha equation corresponding to Eqn. \ref{eqn:Hion}, number conservation, and charge neutrality:
\begin{align}
x_{H,0}&=x_{H,1}\dfrac{n_eZ_{H,0}}{Z_eZ_{H,1}}=x_{H,1}\dfrac{n_e}{n_{\rm q}}\exp\left(\dfrac{T_{\rm ion}}{T}\right)\left(\dfrac{T_{\rm ion}}{T}\right)^{3/2},\\
x_{H,0}&=1-x_{H,1},\\
n_e&=x_{H,1}n_{\rm nuc},
\end{align}
respectively, where $x_{i,r}$ is the fraction of species $i$ in ionization state $r$, and $n_{\rm nuc}=\rho/m_p$ is number density of atomic nuclei regardless of ionization state. Note that we have neglected the electrons' contribution to the mass budget.
Solving these equations for the ionization fraction $x=x_{H,1}$ we get:
\begin{align}
x(\rho,T)&=\left.2\middle/\left(1+\sqrt{1+4\dfrac{\rho}{m_pn_q}\exp\left(\dfrac{T_{\rm ion}}{T}\right)\left(\dfrac{T_{\rm ion}}{T}\right)^{3/2}}\right)\right.\\
x(\trho,\tT)&=\left.2\middle/\left(1+\sqrt{1+4\trho\exp\left(1/\tT\right)\tT^{-3/2}}\right)\right. .
\end{align}
where $\trho=\rho/m_pn_{\rm q}$ and $\tT=T/T_{\rm ion}$.

Pressure and specific internal energy are respectively
\begin{align}
\label{eqn:p}
p&=\sum_{i,r} n_{i,r} kT=n_q k T_{\rm ion}\times \trho \tT \left(1+x(\trho,\tT)\right),\\
\epsilon&=\dfrac{1}{\rho}\sum_{i,r}n_{i,r} kT\der{\ln Z_{i,r}}{\ln T}=\dfrac{k T_{\rm ion}}{m_p}\times\left(x(\trho,\tT)+\dfrac{3}{2}\dfrac{\tP}{\trho}\right).
\end{align}
The generalized adiabatic index $\Gamma_1$ is \citep{PHD}:
\begin{align}
\Gamma_1&=\pder{\ln p}{\ln \rho}{s}=\pder{\ln p}{\ln \rho}{T}+\left[\dfrac{1}{\rho}-\dfrac{\rho}{p}\pder{\epsilon}{\rho}{T}\right]\left.\pder{p}{T}{\rho}\middle/\pder{\epsilon}{T}{\rho}\right.\\
&=\dfrac{5}{3}\left[1 + \left( \tT+\dfrac{2}{3}\right) \left(\dfrac{x_{\tT}}{1+x}\right)\right]^{-1}
\!+\dfrac{5}{3}\dfrac{\left[\frac{4}{15}+\tT\left(\tT+\frac{4}{3}\right)\right]x_{\tT}}{\left(\tT+\frac{2}{3}\right)\left[1+x+\left(\tT+\frac{2}{3}\right)x_{\tT}\right]}
\end{align}
where
\begin{align}
x_{\tT}=\pder{x}{\tT}=\dfrac{x^3}{2-x}\exp\left(\dfrac{1}{\tT}\right)\tT^{-7/2}\left(1+\frac{3}{2}\tT\right)\trho,
\end{align}
making the adiabatic sound speed squared
\begin{align}
a^2=\Gamma_1\dfrac{p}{\rho}=\dfrac{k T_{\rm ion}}{m_p}\times\dfrac{\tP}{\trho}\Gamma_1.
\end{align}

\section{Test Solutions}
\label{sec:sol}

In this appendix, we list all the test errors and all the constant states in the Riemann solutions for all the tests presented in this work. Machine readable tables consisting of a 101 points of data for each rarefaction wave are provided in the supplementary data.

\begin{deluxetable}{cccccccc}
	\tablecaption{Hydrogen HD Test Errors\label{tab:err}}
	\tablehead{
		\colhead{Test} & \colhead{N} & \colhead{$L^1\left(\rho\right)$} & \colhead{$L^2\left(\rho\right)$} & \colhead{$L^1\left(p\right)$} & \colhead{$L^2\left(p\right)$} & \colhead{$L^1\left(v_x\right)$} & \colhead{$L^2\left(v_x\right)$}
	}
	\startdata
	1 & 64 & 1.16821e-09 & 3.91268e-10 & 1.7164e-10 & 4.65662e-11 & 0.00997146 & 0.00494527\\
1 & 128 & 6.43063e-10 & 1.81863e-10 & 8.80361e-11 & 2.15674e-11 & 0.00523906 & 0.00241092\\
1 & 256 & 3.40794e-10 & 9.24717e-11 & 4.3002e-11 & 9.67599e-12 & 0.00257841 & 0.00117046\\
1 & 512 & 1.86891e-10 & 4.80286e-11 & 2.15908e-11 & 4.55806e-12 & 0.0012741 & 0.000559422\\
1 & 1024 & 1.01221e-10 & 2.50378e-11 & 1.09719e-11 & 2.20662e-12 & 0.000633292 & 0.000264448\\
1 & 2048 & 5.48822e-11 & 1.30583e-11 & 5.62261e-12 & 1.11541e-12 & 0.000312803 & 0.000121172\\
\hline
2 & 64 & 1.89475e-08 & 4.34679e-09 & 3.7741e-09 & 9.44887e-10 & 0.0165599 & 0.00908214\\
2 & 128 & 1.0667e-08 & 2.04002e-09 & 1.8781e-09 & 3.92666e-10 & 0.0113663 & 0.00636363\\
2 & 256 & 5.4488e-09 & 8.41316e-10 & 9.41412e-10 & 1.6144e-10 & 0.00436894 & 0.00216652\\
2 & 512 & 2.82345e-09 & 4.80531e-10 & 4.82877e-10 & 7.50044e-11 & 0.00326594 & 0.00183036\\
2 & 1024 & 1.37487e-09 & 1.73961e-10 & 2.33038e-10 & 2.73239e-11 & 0.00121008 & 0.000640447\\
2 & 2048 & 6.92157e-10 & 8.43894e-11 & 1.16744e-10 & 1.21976e-11 & 0.000564997 & 0.000286236\\
\hline
3 & 64 & 1.23061e-07 & 4.57282e-08 & 1.2906e-08 & 6.05245e-09 & 0.0312704 & 0.0192061\\
3 & 128 & 8.57294e-08 & 3.16986e-08 & 1.04151e-08 & 5.81166e-09 & 0.0183161 & 0.0126084\\
3 & 256 & 4.5824e-08 & 1.24077e-08 & 4.02704e-09 & 1.77242e-09 & 0.00622843 & 0.00407155\\
3 & 512 & 3.64848e-08 & 9.45425e-09 & 3.25802e-09 & 1.67845e-09 & 0.00495847 & 0.0033189\\
3 & 1024 & 3.58093e-08 & 8.6303e-09 & 4.4333e-09 & 1.96845e-09 & 0.00640567 & 0.00284412\\
3 & 2048 & 3.0004e-08 & 5.33168e-09 & 3.91971e-09 & 1.37219e-09 & 0.00536005 & 0.00190932\\
\hline
4 & 64 & 1.10813e-07 & 4.92723e-08 & 2.29025e-08 & 1.33251e-08 & 0.0283166 & 0.0218614\\
4 & 128 & 6.09733e-08 & 2.02556e-08 & 9.71378e-09 & 5.79892e-09 & 0.0201343 & 0.0134162\\
4 & 256 & 4.93968e-08 & 1.47944e-08 & 8.3032e-09 & 5.06535e-09 & 0.0162961 & 0.00880635\\
4 & 512 & 4.4765e-08 & 1.08197e-08 & 8.22814e-09 & 3.93867e-09 & 0.0127139 & 0.00603949\\
4 & 1024 & 4.51673e-08 & 8.85295e-09 & 8.48826e-09 & 3.0376e-09 & 0.0114196 & 0.00410761\\
4 & 2048 & 4.41685e-08 & 6.42743e-09 & 8.54673e-09 & 2.21272e-09 & 0.0108255 & 0.00288049\\
\hline
5 & 64 & 1.20788e-06 & 2.2944e-07 & 1.53514e-07 & 2.90521e-08 & 0.0168462 & 0.00285225\\
5 & 128 & 6.45952e-07 & 8.95611e-08 & 7.5403e-08 & 1.08215e-08 & 0.00804194 & 0.00103671\\
5 & 256 & 3.77721e-07 & 4.07579e-08 & 3.77756e-08 & 4.32156e-09 & 0.00601241 & 0.000586455\\
5 & 512 & 2.47277e-07 & 2.10152e-08 & 1.86975e-08 & 1.68397e-09 & 0.00377254 & 0.000313802\\
5 & 1024 & 1.42883e-07 & 1.05143e-08 & 9.2662e-09 & 6.67874e-10 & 0.0020107 & 0.000145441\\
5 & 2048 & 7.29378e-08 & 4.96159e-09 & 4.62453e-09 & 2.60408e-10 & 0.00100041 & 6.327e-05\\
\hline
6 & 64 & 8.91597e-07 & 1.74263e-07 & 1.12256e-07 & 2.20323e-08 & 0.0121635 & 0.00220044\\
6 & 128 & 4.97919e-07 & 7.49812e-08 & 5.63607e-08 & 9.02126e-09 & 0.00637992 & 0.000866221\\
6 & 256 & 2.9138e-07 & 3.27782e-08 & 2.75008e-08 & 3.44633e-09 & 0.00402484 & 0.000409148\\
6 & 512 & 1.74909e-07 & 1.59544e-08 & 1.35664e-08 & 1.3804e-09 & 0.00222649 & 0.000186517\\
6 & 1024 & 9.09769e-08 & 7.14082e-09 & 6.67695e-09 & 5.34788e-10 & 0.00112283 & 8.36673e-05\\
6 & 2048 & 4.58085e-08 & 3.38869e-09 & 3.32678e-09 & 2.1004e-10 & 0.000560512 & 3.70192e-05\\
	\enddata
	\tablecomments{All units are those listed in Table~\ref{tab:units}.}
\end{deluxetable}

\begin{deluxetable}{ccccccc}[!h]
	\tablecaption{Test 1 solution}
	\tablehead{
		\colhead{State ($i$)} & \colhead{$\rho_i$} & \colhead{$p_i$} & \colhead{$v_i$} & \colhead{$T_i$} & \colhead{$\lambda_{\rm min}$} & \colhead{$\lambda_{\rm max}$}
	}
	\startdata
1 & 1.0000000e-07 & 2.9979766e-08 & 0.0000000e+00 & 1.5000000e-01 & $-\infty$ & -7.0412538e-01\\
2 & 3.6231794e-08 & 6.5530353e-09 & 5.9219500e-01 & 9.2937185e-02 & 1.0725946e-01 & 5.9219500e-01\\
3 & 5.9466421e-08 & 6.5530353e-09 & 5.9219500e-01 & 7.4070032e-02 & 5.9219500e-01 & 7.4980628e-01\\
4 & 1.2500000e-08 & 1.0026412e-09 & 0.0000000e+00 & 6.2000000e-02 & 7.4980628e-01 & $\infty$\\

	\enddata
	\tablecomments{Density ($\rho$), pressure ($p$), speed ($u$), temperature ($T$) and bounding wave speeds ($\lambda_{\rm min}, \lambda_{\rm max}$) for the four constant states in the solution of test 1. The solution for the rarefaction wave is in supplementary material.}
\end{deluxetable}

\begin{deluxetable}{ccccccc}
	\tablecaption{Test 2 solution}
	\tablehead{
		\colhead{State ($i$)} & \colhead{$\rho_i$} & \colhead{$p_i$} & \colhead{$v_i$} & \colhead{$T_i$} & \colhead{$\lambda_{\rm min}$} & \colhead{$\lambda_{\rm max}$}
	}
	\startdata
1 & 4.0000000e-06 & 8.4741487e-07 & 0.0000000e+00 & 1.2000000e-01 & $-\infty$ & -5.1742727e-01\\
2 & 3.8242193e-07 & 4.9053229e-08 & 1.0452190e+00 & 8.5178414e-02 & 6.5617334e-01 & 1.0452190e+00\\
3 & 4.2049684e-07 & 4.9053229e-08 & 1.0452190e+00 & 8.2328850e-02 & 1.0452190e+00 & 1.1550984e+00\\
4 & 4.0000000e-08 & 7.6000000e-10 & 0.0000000e+00 & 1.9000000e-02 & 1.1550984e+00 & $\infty$\\

	\enddata
	\tablecomments{Density ($\rho$), pressure ($p$), speed ($u$), temperature ($T$) and bounding wave speeds ($\lambda_{\rm min}, \lambda_{\rm max}$) for the four constant states in the solution of test 2. The solution for the rarefaction wave is in supplementary material.}
\end{deluxetable}

\begin{deluxetable}{ccccccc}
	\tablecaption{Test 3 solution}
	\tablehead{
		\colhead{State ($i$)} & \colhead{$\rho_i$} & \colhead{$p_i$} & \colhead{$v_i$} & \colhead{$T_i$} & \colhead{$\lambda_{\rm min}$} & \colhead{$\lambda_{\rm max}$}
	}
	\startdata
1 & 8.0000000e-07 & 4.8000000e-09 & 1.1000000e+00 & 6.0000000e-03 & $-\infty$ & -1.8903927e-01\\
2 & 7.7533506e-06 & 1.1969393e-06 & -5.6034656e-02 & 1.0639183e-01 & -1.8903927e-01 & -5.6034656e-02\\
3 & 4.2101848e-06 & 1.1969393e-06 & -5.6034656e-02 & 1.4659754e-01 & -5.6034656e-02 & 1.1655176e-01\\
4 & 4.0000000e-07 & 2.4000000e-09 & -1.7000000e+00 & 6.0000000e-03 & 1.1655176e-01 & $\infty$\\

	\enddata
	\tablecomments{Density ($\rho$), pressure ($p$), speed ($u$), temperature ($T$) and bounding wave speeds ($\lambda_{\rm min}, \lambda_{\rm max}$) for the four constant states in the solution of test 3.}
\end{deluxetable}

\begin{deluxetable}{ccccccc}
	\tablecaption{Test 4 solution}
	\tablehead{
		\colhead{State ($i$)} & \colhead{$\rho_i$} & \colhead{$p_i$} & \colhead{$v_i$} & \colhead{$T_i$} & \colhead{$\lambda_{\rm min}$} & \colhead{$\lambda_{\rm max}$}
	}
	\startdata
1 & 5.0000000e-07 & 3.0000000e-09 & 1.5000000e+00 & 6.0000000e-03 & $-\infty$ & -2.2343227e-01\\
2 & 5.3963971e-06 & 1.3505075e-06 & -6.3748668e-02 & 1.3442604e-01 & -2.2343227e-01 & -6.3748668e-02\\
3 & 3.7900129e-06 & 1.3505075e-06 & -6.3748668e-02 & 1.7931867e-01 & -6.3748668e-02 & 1.4111796e-01\\
4 & 4.0000000e-07 & 2.4000000e-09 & -1.8000000e+00 & 6.0000000e-03 & 1.4111796e-01 & $\infty$\\

	\enddata
	\tablecomments{Density ($\rho$), pressure ($p$), speed ($u$), temperature ($T$) and bounding wave speeds ($\lambda_{\rm min}, \lambda_{\rm max}$) for the four constant states in the solution of test 4.}
\end{deluxetable}

\begin{deluxetable}{ccccccc}
	\tablecaption{Test 5 solution}
	\tablehead{
		\colhead{State ($i$)} & \colhead{$\rho_i$} & \colhead{$p_i$} & \colhead{$v_i$} & \colhead{$T_i$} & \colhead{$\lambda_{\rm min}$} & \colhead{$\lambda_{\rm max}$}
	}
	\startdata
1 & 8.0000000e-05 & 8.3166294e-06 & -8.0000000e-01 & 9.5000000e-02 & $-\infty$ & -1.1617972e+00\\
2 & 6.1125432e-06 & 2.5241908e-07 & 0.0000000e+00 & 4.1286848e-02 & -2.6033771e-01 & 0.0000000e+00\\
3 & 6.1125432e-06 & 2.5241908e-07 & 0.0000000e+00 & 4.1286848e-02 & 0.0000000e+00 & 2.6033771e-01\\
4 & 8.0000000e-05 & 8.3166294e-06 & 8.0000000e-01 & 9.5000000e-02 & 1.1617972e+00 & $\infty$\\

	\enddata
	\tablecomments{Density ($\rho$), pressure ($p$), speed ($u$), temperature ($T$) and bounding wave speeds ($\lambda_{\rm min}, \lambda_{\rm max}$) for the four constant states in the solution of test 5. The solutions for the rarefaction waves are in supplementary material.}
\end{deluxetable}

\begin{deluxetable}{ccccccc}
	\tablecaption{Test 6 solution}
	\tablehead{
		\colhead{State ($i$)} & \colhead{$\rho_i$} & \colhead{$p_i$} & \colhead{$v_i$} & \colhead{$T_i$} & \colhead{$\lambda_{\rm min}$} & \colhead{$\lambda_{\rm max}$}
	}
	\startdata
1 & 6.0000000e-05 & 6.3158878e-06 & -5.0000000e-01 & 9.5000000e-02 & $-\infty$ & -8.6273248e-01\\
2 & 7.1322370e-06 & 4.1186956e-07 & 1.8235310e-01 & 5.7338347e-02 & -1.0278163e-01 & 1.8235310e-01\\
3 & 8.2935436e-06 & 4.1186956e-07 & 1.8235310e-01 & 4.9585965e-02 & 1.8235310e-01 & 4.6030997e-01\\
4 & 8.0000000e-05 & 8.3166294e-06 & 9.0000000e-01 & 9.5000000e-02 & 1.2617972e+00 & $\infty$\\

	\enddata
	\tablecomments{Density ($\rho$), pressure ($p$), speed ($u$), temperature ($T$) and bounding wave speeds ($\lambda_{\rm min}, \lambda_{\rm max}$) for the four constant states in the solution of test 6. The solutions for the rarefactions wave are in supplementary material.}
\end{deluxetable}

\clearpage
\begin{turnpage}
\begin{deluxetable}{ccccccccccccccc}
	\tablecaption{Test 7 (Hydrogen MHD) Errors\label{tab:mhd_err}}
	\tablehead{
		\colhead{N} & \colhead{$L^1\left(\rho\right)$} & \colhead{$L^2\left(\rho\right)$} & \colhead{$L^1\left(v_x\right)$} & \colhead{$L^2\left(v_x\right)$} & \colhead{$L^1\left(v_y\right)$} & \colhead{$L^2\left(v_y\right)$} & \colhead{$L^1\left(v_z\right)$} & \colhead{$L^2\left(v_z\right)$} & \colhead{$L^1\left(B_y\right)$} & \colhead{$L^2\left(B_y\right)$} & \colhead{$L^1\left(B_z\right)$} & \colhead{$L^2\left(B_z\right)$} & \colhead{$L^1\left(p\right)$} & \colhead{$L^2\left(p\right)$}
	}
	\startdata
	64 & 2.07e-2 & 5.44e-3 & 1.47e-2 & 5.42e-3 & 1.18e-2 & 3.31e-3 & 8.20e-3 & 2.48e-3 & 1.78e-2 & 5.06e-3 & 1.37e-2 & 3.71e-3 & 2.23e-2 & 6.48e-3\\
128 & 1.17e-2 & 2.72e-3 & 7.76e-3 & 2.74e-3 & 7.09e-3 & 1.72e-3 & 5.44e-3 & 1.52e-3 & 1.06e-2 & 2.54e-3 & 8.16e-3 & 2.04e-3 & 1.22e-2 & 3.19e-3\\
256 & 6.27e-3 & 1.33e-3 & 3.90e-3 & 1.32e-3 & 3.73e-3 & 8.62e-4 & 3.56e-3 & 8.63e-4 & 5.47e-3 & 1.21e-3 & 4.92e-3 & 1.11e-3 & 6.28e-3 & 1.59e-3\\
512 & 3.32e-3 & 7.21e-4 & 2.09e-3 & 7.66e-4 & 2.05e-3 & 4.29e-4 & 2.06e-3 & 4.95e-4 & 3.01e-3 & 6.80e-4 & 2.84e-3 & 6.51e-4 & 3.27e-3 & 8.53e-4\\
1024 & 1.77e-3 & 3.59e-4 & 9.76e-4 & 3.17e-4 & 1.11e-3 & 2.16e-4 & 1.16e-3 & 2.55e-4 & 1.56e-3 & 3.16e-4 & 1.56e-3 & 3.26e-4 & 1.62e-3 & 3.96e-4\\
2048 & 9.85e-4 & 1.93e-4 & 5.03e-4 & 1.78e-4 & 6.20e-4 & 1.16e-4 & 6.97e-4 & 1.43e-4 & 8.83e-4 & 1.71e-4 & 9.30e-4 & 1.82e-4 & 8.66e-4 & 2.07e-4\\
	\enddata
	\tabletypesize{\normalsize}
	\tablecomments{Errors for test 7. Density is in units of $10^{-7}\rho_{\rm u}$, velocity in $\sqrt{0.2}\,v_{\rm u}$, magnetic fields in $\sqrt{2\times 10^{-8}} \,B_{\rm u}$, pressure in $2\times 10^{-8} \,p_{\rm u}$ (see Table~\ref{tab:units} for unit deffinitions).}
\end{deluxetable}

\begin{deluxetable}{cccccccccc}
	\tablecaption{Test 7 solution}
	\tablehead{
		\colhead{State ($i$)} & \colhead{$\rho$} & \colhead{$v_x$} & \colhead{$v_y$} & \colhead{$v_z$} & \colhead{$B_y$} & \colhead{$B_z$} & \colhead{$p$} & \colhead{$\lambda_{\rm min}$} & \colhead{$\lambda_{\rm max}$}
	}
	\tabletypesize{\footnotesize}
	\startdata
1 & 1.0800000e+00 & 1.2000000e+00 & 1.0000000e-02 & 5.0000000e-01 & 1.0155413e+00 & 5.6418958e-01 & 9.5000000e-01 & $-\infty$ & -8.2366304e-01\\
2 & 1.5272580e+00 & 6.0736964e-01 & 1.3086752e-01 & 5.6714862e-01 & 1.4837575e+00 & 8.2430973e-01 & 1.4795324e+00 & -8.2366304e-01 & 1.5084013e-01\\
3 & 1.5272580e+00 & 6.0736964e-01 & 2.4201900e-01 & 3.0713627e-01 & 1.6211210e+00 & 5.0298054e-01 & 1.4795324e+00 & 1.5084013e-01 & 2.9445282e-01\\
4 & 1.7756066e+00 & 5.6360293e-01 & 3.0981899e-02 & 2.4165840e-01 & 1.4423591e+00 & 4.4751659e-01 & 1.8006247e+00 & 2.9445282e-01 & 5.6360293e-01\\
5 & 1.5041659e+00 & 5.6360293e-01 & 3.0981899e-02 & 2.4165840e-01 & 1.4423591e+00 & 4.4751659e-01 & 1.8006247e+00 & 5.6360293e-01 & 8.6856879e-01\\
6 & 1.3165385e+00 & 5.2014049e-01 & -1.9219053e-01 & 1.7241534e-01 & 1.6238114e+00 & 5.0381526e-01 & 1.4757265e+00 & 8.6856879e-01 & 1.0118496e+00\\
7 & 1.3165385e+00 & 5.2014049e-01 & -1.0231023e-01 & -5.1155117e-02 & 1.5206823e+00 & 7.6034114e-01 & 1.4757265e+00 & 1.0118496e+00 & 2.1633547e+00\\
8 & 1.0000000e+00 & 0.0000000e+00 & 0.0000000e+00 & 0.0000000e+00 & 1.1283792e+00 & 5.6418958e-01 & 1.0000000e+00 & 2.1633547e+00 & $\infty$\\

	\enddata
	\tabletypesize{\normalsize}
	\tablecomments{Density ($\rho$), velocity ($v_{x,y,z}$), magnetic field ($B_{y,z}$), pressure ($p$) and bounding wave speeds ($\lambda_{\rm min}, \lambda_{\rm max}$) for the eight constant states in the solution of test 7. For this test $B_x=2/\sqrt{4\pi}$ is constant accross all states.}
	\label{tab:sol-7}
\end{deluxetable}
\end{turnpage}

\begin{deluxetable}{ccccccc}
	\tablecaption{Test 8 (HD Linear Wave) Errors\label{tab:hd_lin-wave_err}}
	\tablehead{
		\colhead{N} & \colhead{$L^1\left(\rho\right)$} & \colhead{$L^2\left(\rho\right)$} & \colhead{$L^1\left(\rho v_x\right)$} & \colhead{$L^2\left(\rho v_x\right)$} & \colhead{$L^1\left(E\right)$} & \colhead{$L^2\left(E\right)$}
	}
	\vspace*{.5em}
	\startdata
	64 & 6.36583e-16 & 9.83326e-16 & 1.59797e-16 & 2.46838e-16 & 7.13911e-17 & 1.10277e-16\\
128 & 1.46028e-16 & 2.84331e-16 & 3.66563e-17 & 7.13736e-17 & 1.63766e-17 & 3.18869e-17\\
256 & 3.32521e-17 & 8.11102e-17 & 8.34705e-18 & 2.03606e-17 & 3.72914e-18 & 9.0963e-18\\
512 & 7.46361e-18 & 2.297e-17 & 1.87354e-18 & 5.76602e-18 & 8.37023e-19 & 2.57603e-18\\
1024 & 1.7267e-18 & 6.46748e-18 & 4.33441e-19 & 1.62349e-18 & 1.93644e-19 & 7.25312e-19\\
2048 & 5.30243e-19 & 1.82464e-18 & 1.33104e-19 & 4.58027e-19 & 5.94659e-20 & 2.04629e-19\\
	\enddata
	\tablecomments{Errors for test 8. Units are those defined in Table~\ref{tab:units}.}
\end{deluxetable}

\begin{deluxetable}{ccccccccc}
	\tablecaption{Test 9 (HD Linear Wave)
	 Errors\label{tab:mhd_lin-wave_err}}
	\tablehead{
		\colhead{N} & \colhead{$L^1\left(\rho\right)$} & \colhead{$L^2\left(\rho\right)$} &  \colhead{$L^1\left(E\right)$} & \colhead{$L^2\left(E\right)$} & \colhead{$L^1\left(B_y\right)$} & \colhead{$L^2\left(B_y\right)$} & \colhead{$L^1\left(B_z\right)$} & \colhead{$L^2\left(B_z\right)$}
	}
	\vspace*{.5em}
	\startdata
	64 & 6.29632e-16 & 9.89452e-16 & 1.51959e-16 & 2.35843e-16 & 7.74981e-13 & 1.18933e-12 & 2.73997e-13 & 4.20492e-13\\
128 & 1.44955e-16 & 2.8507e-16 & 3.48544e-17 & 6.81348e-17 & 1.78191e-13 & 3.44357e-13 & 6.30001e-14 & 1.21748e-13\\
256 & 3.29876e-17 & 8.12143e-17 & 7.94064e-18 & 1.94313e-17 & 4.06191e-14 & 9.82814e-14 & 1.4361e-14 & 3.47477e-14\\
512 & 7.41914e-18 & 2.29877e-17 & 1.77981e-18 & 5.50238e-18 & 9.15496e-15 & 2.7838e-14 & 3.2369e-15 & 9.84223e-15\\
1024 & 1.77304e-18 & 6.47438e-18 & 4.2569e-19 & 1.5499e-18 & 2.16908e-15 & 7.8419e-15 & 7.66903e-16 & 2.77256e-15\\
2048 & 6.02096e-19 & 1.83859e-18 & 1.42777e-19 & 4.39714e-19 & 7.22467e-16 & 2.22464e-15 & 2.55455e-16 & 7.86526e-16\\
	\enddata
\end{deluxetable}

\begin{deluxetable}{ccccccc}
	\vspace*{-1.1in}
	{\hfill Table 15 Continued \hfill}
	\tablehead{
	\colhead{N} & 
	\colhead{$L^1\left(\rho v_x\right)$} &
	\colhead{$L^2\left(\rho v_x\right)$} &
	\colhead{$L^1\left(\rho v_y\right)$} & 
	\colhead{$L^2\left(\rho v_y\right)$} & 
	\colhead{$L^1\left(\rho v_z\right)$} & 
	\colhead{$L^2\left(\rho v_z\right)$}
}
\vspace*{.5em}
\startdata
64 & 2.80339e-16 & 4.29146e-16 & 1.14097e-16 & 1.80366e-16 & 4.03395e-17 & 6.37689e-17\\
128 & 6.44171e-17 & 1.24093e-16 & 2.60402e-17 & 5.21443e-17 & 9.2066e-18 & 1.84358e-17\\
256 & 1.47207e-17 & 3.53998e-17 & 5.8737e-18 & 1.48757e-17 & 2.07667e-18 & 5.25935e-18\\
512 & 3.31368e-18 & 1.00249e-17 & 1.31216e-18 & 4.21307e-18 & 4.63906e-19 & 1.48954e-18\\
1024 & 7.83211e-19 & 2.82383e-18 & 3.20466e-19 & 1.18671e-18 & 1.13301e-19 & 4.19567e-19\\
2048 & 2.61529e-19 & 8.01252e-19 & 1.06154e-19 & 3.36192e-19 & 3.75288e-20 & 1.18864e-19\\
\enddata
\tablecomments{Errors for test 9. Units are those defined in Table~\ref{tab:units}.}
\end{deluxetable}

\begin{deluxetable}{cccccccc}
	\tablecaption{Helmholtz HD Test 10 Errors\label{tab:helm_err}}
	\tablehead{
		\colhead{Test} & \colhead{N} & \colhead{$L^1\left(\rho\right)$} & \colhead{$L^2\left(\rho\right)$} & \colhead{$\dfrac{L^1\left(p\right)}{10^{16}}$} & \colhead{$\dfrac{L^2\left(p\right)}{10^{16}}$} & \colhead{$\dfrac{L^1\left(v_x\right)}{10^8}$} & \colhead{$\dfrac{L^2\left(v_x\right)}{10^8}$}
		\vspace*{.5em}
	}
	\startdata
	10 & 64 & 0.00842 & 0.00214067 & 0.004111 & 0.00114166 & 0.0129963 & 0.00536057\\
10 & 128 & 0.00411609 & 0.00120448 & 0.00206671 & 0.000564817 & 0.00806605 & 0.00432055\\
10 & 256 & 0.00210724 & 0.000637162 & 0.00103097 & 0.000284614 & 0.00420335 & 0.00230794\\
10 & 512 & 0.00116523 & 0.000364281 & 0.000532187 & 0.000157629 & 0.00227806 & 0.00128378\\
10 & 1024 & 0.000515071 & 0.000122224 & 0.000243799 & 5.5439e-05 & 0.000799265 & 0.00033411\\
10 & 2048 & 0.000282591 & 7.7422e-05 & 0.000125973 & 3.23936e-05 & 0.000513506 & 0.000280413\\
\hline
11 & 64 & 0.0455309 & 0.0159875 & 0.0206146 & 0.00774235 & 0.0229773 & 0.0119659\\
11 & 128 & 0.0254998 & 0.00867462 & 0.0122055 & 0.00477635 & 0.0111093 & 0.00634357\\
11 & 256 & 0.0147393 & 0.00562628 & 0.00684821 & 0.00317944 & 0.00599355 & 0.00351935\\
11 & 512 & 0.00677742 & 0.0020821 & 0.00286122 & 0.00115752 & 0.00202944 & 0.000870735\\
11 & 1024 & 0.00395334 & 0.00129364 & 0.00155555 & 0.000733798 & 0.00173115 & 0.00114641\\
11 & 2048 & 0.00222715 & 0.000784014 & 0.00085564 & 0.000421076 & 0.000749624 & 0.000435848\\
	\enddata
	\tablecomments{Values given in cgs units.}
\end{deluxetable}

\begin{deluxetable}{ccccccc}
	\tablecaption{Helmholtz HD Test 10 solution}
	\tablehead{
		\colhead{State ($i$)} & \colhead{$\rho_i$} & \colhead{$\dfrac{p_i}{10^{16}}$} & \colhead{$\dfrac{v_i}{10^{8}}$} & \colhead{$T_i$} & \colhead{$\dfrac{\lambda_{\rm min}}{10^{8}}$} & \colhead{$\dfrac{\lambda_{\rm max}}{10^{8}}$}
		\vspace*{.5em}
	}
	\startdata
1 & 1.0000000e+00 & 7.0000000e-01 & 0.0000000e+00 & 2.9928464e+07 & $-\infty$ & -1.0173409e+00\\
2 & 3.6995110e-01 & 1.5962459e-01 & 9.0151646e-01 & 1.9732827e+07 & 9.8304277e-02 & 9.0151646e-01\\
3 & 4.2009694e-01 & 1.5962459e-01 & 9.0151646e-01 & 1.8563834e+07 & 9.0151646e-01 & 1.2833895e+00\\
4 & 1.2500000e-01 & 1.5000000e-02 & 0.0000000e+00 & 6.9368772e+06 & 1.2833895e+00 & $\infty$\\

	\enddata
	\tablecomments{Density ($\rho$), pressure ($p$), speed ($u$), temperature ($T$) and bounding wave speeds ($\lambda_{\rm min}, \lambda_{\rm max}$) in cgs units for the four constant states in the solution of test 10. The solutions for the rarefactions wave are in supplementary material.}
\end{deluxetable}

\begin{deluxetable}{ccccccc}
	\tablecaption{Test 11 solution}
	\tablehead{
		\colhead{State ($i$)} & \colhead{$\rho_i$} & \colhead{$\dfrac{p_i}{10^{16}}$} & \colhead{$\dfrac{v_i}{10^{8}}$} & \colhead{$T_i$} & \colhead{$\dfrac{\lambda_{\rm min}}{10^{8}}$} & \colhead{$\dfrac{\lambda_{\rm max}}{10^{8}}$}
		\vspace*{.5em}
	}
	\startdata
1 & 8.0000000e-01 & 5.0000000e-02 & 1.1000000e+00 & 3.7568121e+06 & $-\infty$ & -4.4550675e-01\\
2 & 3.0224360e+00 & 1.4550894e+00 & -3.6430963e-02 & 2.6482111e+07 & -4.4550675e-01 & -3.6430963e-02\\
3 & 1.7919758e+00 & 1.4550894e+00 & -3.6430963e-02 & 3.5455933e+07 & -3.6430963e-02 & 4.4161443e-01\\
4 & 4.0000000e-01 & 3.0000000e-02 & -1.7000000e+00 & 4.4969785e+06 & 4.4161443e-01 & $\infty$\\

	\enddata
	\tablecomments{Density ($\rho$), pressure ($p$), speed ($u$), temperature ($T$) and bounding wave speeds ($\lambda_{\rm min}, \lambda_{\rm max}$) in cgs units for the four constant states in the solution of test 11.}
\end{deluxetable}

\end{document}